\documentclass[useAMS,usenatbib]{mn2e}

\usepackage{graphicx}
\usepackage[figuresright]{rotating}

\newcommand{\kms}{\textrm{km\,s$^{-1}$}}

\newcommand{\Hbeta}{H$\beta$}
\newcommand{\HgA}{H$\gamma_A$}
\newcommand{\HgF}{H$\gamma_F$}
\newcommand{\HdA}{H$\delta_A$}
\newcommand{\HdF}{H$\delta_F$}
\newcommand{\Fe}{$\langle${Fe}$\rangle$}
\newcommand{\ZMgZFe}{[Z$_{{\rm Mg}b}$/Z$_{\langle{\rm Fe}\rangle}$]}

\title[dEs from the MAGPOP-ITP]{The relation between stellar populations,
  structure and environment for dwarf elliptical galaxies from the
  MAGPOP-ITP.}

\author[D. Michielsen et al.]{%
  D. Michielsen$^1$,
  A. Boselli$^2$, 
  C. J. Conselice$^1$, 
  E. Toloba$^3$, 
  I.~M. Whiley$^1$,   \newauthor
  A. Arag\'on-Salamanca$^1$, 
  M. Balcells$^4$,
  N. Cardiel$^3$,
  A.~J. Cenarro$^4$, 
  J. Gorgas$^3$, \newauthor
  R.~F. Peletier$^5$,
  A. Vazdekis$^4$
  \\
  $^{1}$ School of Physics and Astronomy, University of Nottingham, University
  Park, Nottingham NG7 2RD, UK\\
  $^{2}$ Laboratoire d'Astrophysique de Marseille, BP8, Traverse du Siphon,
  F-13376 Marseille, France\\
  $^{3}$ Departamento de F\'{i}sica de la Tierra, Astronom\'{i}a y
  Astrof\'{i}sica II, Universidad Complutense de Madrid, E-28040 Madrid,
  Spain\\
  $^{4}$ Instituto de Astrof\'{i}sica de Canarias, V\'{i}a L\'actea s/n,
  E-38200, La Laguna, Tenerife, Spain\\
  $^{5}$ Kapteyn Astronomical Institute, University of Groningen, Postbus 800,
  9700 AV Groningen, the Netherlands
}

\begin{document}

\date{}
\pagerange{\pageref{firstpage}--\pageref{lastpage}} \pubyear{2006}
\maketitle
\label{firstpage}
\begin{abstract}
  Dwarf galaxies, as the most numerous type of galaxy, offer the
  potential to study galaxy formation and evolution in detail in the
  nearby Universe.  Although they seem to be simple systems at first
  view, they remain poorly understood. In an attempt to alleviate this
  situation, the MAGPOP EU Research and Training Network embarked on a
  study of dwarf galaxies named MAGPOP-ITP \citep{peletier07}. In this
  paper, we present the analysis of a sample of 24 dwarf elliptical
  galaxies (dEs) in the Virgo Cluster and in the field, using optical
  long-slit spectroscopy. We examine their stellar populations in
  combination with their light distribution and environment.  We
  confirm and strengthen previous results that dEs are, on average,
  younger and more metal-poor than normal elliptical galaxies, and
  that their [$\alpha$/Fe] abundance ratios scatter around solar.
  This is in accordance with the downsizing picture of galaxy
  formation where mass is the main driver for the star formation
  history. We also find new correlations between the
  luminosity-weighted mean age, the large-scale asymmetry, and the
  projected Virgocentric distance. We find that environment plays an
  important role in the termination of the star formation activity by
  ram pressure stripping of the gas in short timescales, and in the
  transformation of disky dwarfs to more spheroidal objects by
  harassment over longer timescales. This points towards a continuing
  infall scenario for the evolution of dEs.
\end{abstract}


\section{Introduction}
\label{sec_intro}

Dwarf galaxies are the lowest mass galaxies in the universe, and are the most
common galaxy type. Particularly, dwarf elliptical galaxies\footnote{In this
  paper, we use dE for all low-luminosity ($M_B > -18$) early-type galaxies:
  dwarf elliptical, dwarf lenticular and dwarf spheroidal galaxies.} (dEs) are
the dominant galaxy population in galaxy clusters, and their evolved nature
and abundance makes them ideal targets for detailed study
\citep{sandageetal85, fergusonbinggeli94}. Their properties and evolution also
reveal much about galaxy formation in general, and can serve as a test of
cosmological models.  According to the widely accepted $\Lambda$CDM
hierarchical merging scenario, dwarf-size dark matter halos are the first to
form, and higher mass galaxies are thought to form from the merging of these
low-mass systems \citep[e.g.][]{whiterees78, whitefrenk91}. By studying dwarf
galaxies we can therefore potentially study the first galaxies, or at least
those galaxies with very simple formation histories.

However, it appears that, observationally, the star formation in lower mass
galaxies is shut off, or exhausted, later than in giant galaxies or, in other
words, that massive galaxies formed earlier and more quickly
\citep{cowieetal96, gavazzietal96, bosellietal01, caldwelletal03, nelanetal05,
  bundyetal06}.  Although seemingly anti-hierarchical, this behaviour can be
reproduced in semi-analytic simulations of galaxy formation in a $\Lambda$CDM
cosmology \citep{deluciaetal06}. This ``downsizing'' puts the observational
study of dwarf and low-mass galaxies into a new focus. Although some dEs in
clusters have old stellar populations, as seen through their globular cluster
systems \citep{beasleyetal06, conselice06}, it is clear that not all dwarfs
have a single formation event, but appear to form in several star formation
episodes.

Detailed study of dwarfs in groups, especially in the Local Group, reveals
that most dEs have a broad star formation history, with many appearing to have
a star formation burst or event in the past few Gyr \citep[e.g.][]{mateo98,
  grebeletal03}. Nearly all of these Local Group dwarfs also have old stellar
populations that date back to roughly the time of reionisation. These results,
however, are based on the study of resolved stellar populations, and more
distant dwarfs cannot be as easily studied. Based on the velocity distribution
of the dE population as a whole \citep{conseliceetal01}, and on their stellar
populations \citep[e.g.][]{caldwelletal03, poggiantietal01, rakosetal01,
  vanzeeetal04}, evidence accumulated to date suggests that dEs in nearby
clusters have a mixed origin. Some have properties consistent with an old
primordial formation, while others appear to be more recently formed from
accreted field galaxies. It is thus becoming clear that dEs are not just small
Es with simple, old and metal-poor stellar populations.

An independent test of the star formation history of galaxies comes from their
integrated stellar populations and, more specifically, from their
[$\alpha$/Fe] abundance ratios \citep{wortheyetal92}. Since $\alpha$ elements,
such as Mg, are mainly produced on short timescales by type~II supernovae,
while most of the Fe is formed later by type~Ia supernovae, the observed
super-solar [$\alpha$/Fe] abundance ratios in giant ellipticals (Es) is
attributed to short formation timescales. The observed correlation of
[$\alpha$/Fe] abundance ratio and galaxy mass is again a manifestation of the
downsizing \citep{vazdekisetal04,thomasetal05, nelanetal05}.
\citet{gorgasetal97} were the first to recognise that Virgo dEs are consistent
with solar [$\alpha$/Fe] abundance ratios, pointing to a more gradual chemical
evolution in low-mass systems.  Later studies have confirmed these results and
also found that on average, dEs have lower metallicities and younger ages than
normal Es \citep{gehaetal03, vanzeeetal04}.

Scenarios for the formation and evolution of dEs are still actively debated.
On the one hand, internal processes play a role, mainly through supernova
feedback. On the other hand, because of their low masses, the properties of
dwarf galaxies are expected to depend strongly on the environment they reside
in. For example, the morphology-density relation, also observed for massive
galaxies, is indeed very strong for low-mass galaxies
\citep[e.g.][]{binggelietal87}.

Supernova feedback regulates and/or suppresses star formation, eventually
leading to gas exhaustion through star formation and/or to gas expulsion
through galactic winds \citep[e.g.][]{daviesphillips88, carraroetal01,
  dekelwoo03}. Invoking only internal processes can reproduce observed
structural and kinematical correlations for dEs \citep{derijckeetal05}, but of
course not the morphology-density relation. In a dense environment a variety
of external processes act on galaxies and may even transform late-type
galaxies into early-type galaxies. This transformation depends on the
environment and involves several mechanisms. For instance, ram-pressure
stripping by the hot intracluster medium can deprive a galaxy of its gas
\citep{gunngott72}, while harassment by galaxy-galaxy interactions transforms
disks into more spheroidal objects \citep{mooreetal98}. For an extensive
review on these, and other, environmental effects see
\citet{boselligavazzi06}. Observations of rotation in dEs, and the existence
of dEs with residual disk structure, support the idea that some dEs are
transformed late-type spiral or dwarf irregular galaxies \citep{pedrazetal02,
  simienprugniel02, derijckeetal03disks, gehaetal03, vanzeeetal04rotation,
  liskeretal06disk}.

Given the importance of dwarf galaxies and the fact that we still do not know
what mechanisms play the dominant role in their formation and evolution, the
MAGPOP EU Research and Training Network embarked on an observational project
to study the star formation history of dwarf galaxies. In the framework of an
International Time Programme (ITP) we used a variety of telescopes and
instruments at the Roque de los Muchachos Observatory in La Palma to study the
structure, dynamics and stellar populations in a large sample of dwarf
galaxies \citep[(see][]{peletier07}. In this paper we present the first
results of this project, examining the ages, metallicities and abundance
ratios of dEs and their relation to their stellar light distribution and
environment, using intermediate-resolution optical spectra.

In the next section we describe the sample, the observations and the data
reduction. In Section~\ref{sec_results}, we report the results from our
analysis of the indices (\ref{sec_indices}), ages, metallicities and abundance
ratios (\ref{sec_agemetalfa}), the stellar light distributions
(\ref{sec_lightdistribution}) and the Virgocentric distance
(\ref{sec_virgodistance}). We discuss the implications of our results on the
formation and evolution scenarios for dEs in Section~\ref{sec_discussion}.
Finally we summarise our conclusions in Section~\ref{sec_conclusions}.

\section{Sample, observations and data reduction}
\label{sec_sample_obs_datared}

\begin{table*}
\begin{minipage}{170mm} 
  \caption{The sample: basic properties}
  \label{tab_sample}
  \begin{tabular}{llcclcccccc}
    \hline
    \multicolumn{1}{c}{galaxy} & \multicolumn{1}{c}{alt. name}
                       & RA (J2000) & DEC (J2000) & \multicolumn{1}{c}{Type}
                                                       & $r_{{\rm eff},H}$ & $D$  & $B_T$ & $M_B$ & note \\
             &  & (h, m, s) & ($\deg$,\arcmin,\arcsec) &       &(\arcsec)&(Mpc) & (mag) & (mag) &      \\
   \multicolumn{1}{c}{(1)} & \multicolumn{1}{c}{(2)} 
                             &     (3)    &     (4)     &\multicolumn{1}{c}{(5)}
                                                                 &  (6)  & (7)  &  (8)  &  (9)  & (10) \\
    \hline                                                                               
    M\,32     & NGC 221      & 00:42:41.84 & +40:51:57.4 & cE2    &  ---  & 1.97 &  8.89 &$-17.58$&     \\
    ID\,0650  & UGC 8986     & 14:04:15.87 & +04:06:43.9 & S0?    &  ---  & 17.8 & 15.06 &$-16.19$&     \\
    ID\,0734  & PGC 1007217  & 02:41:35.08 & $-$08:10:24.8 & ---  &  ---  & 22.0 & 15.92 &$-15.79$&     \\
    ID\,0872  & PGC 1154903  & 02:42:00.37 & +00:00:52.3 & ---    &  ---  & 16.0 & 17.03 &$-13.99$&     \\
    ID\,0918  & CGCG 020-039 & 14:58:48.76 & +02:01:24.9 & E      &  ---  & 25.9 & 14.81 &$-17.26$&     \\
    ID\,1524  & NGC 5870     & 15:06:33.86 & +55:28:46.0 & S0?    &  ---  & 11.8 & 14.83 &$-15.53$&     \\
    VCC\,0021 & IC 3025      & 12:10:23.14 & +10:11:18.9 & dS0(4) & 11.95 & 17.0 & 14.91 &$-16.24$&     \\
    VCC\,0308 & IC 3131/3132 & 12:18:50.77 & +07:51:41.3 & dS0,N: & 17.11 & 23.0 & 14.42 &$-17.39$&     \\
    VCC\,0397 & CGCG 042-031 & 12:20:12.25 & +06:37:23.6 & dE5,N  &  ---  & 23.0 & 15.18 &$-16.63$&     \\ 
    VCC\,0523 & NGC 4306     & 12:22:04.13 & +12:47:15.1 & dSB0,N & 17.90 & 17.0 & 13.81 &$-17.34$&     \\ 
    VCC\,0856 & IC 3328      & 12:25:57.93 & +10:03:13.8 & dE1,N  & 14.32 & 23.0 & 14.48 &$-17.33$& 1   \\ 
    VCC\,0917 & IC 3344      & 12:26:32.40 & +13:34:43.8 & dE6    &  9.29 & 17.0 & 15.45 &$-15.70$& 1,2 \\ 
    VCC\,0990 & IC 3369      & 12:27:16.91 & +16:01:28.4 & dE4,N  &  ---  & 17.0 & 14.88 &$-16.27$& 2   \\ 
    VCC\,1087 & IC 3381      & 12:28:14.88 & +11:47:23.7 & dE3,N  & 15.93 & 17.0 & 14.38 &$-16.77$& 1   \\ 
    VCC\,1122 & IC 3393      & 12:28:41.74 & +12:54:57.3 & dE7,N  & 11.82 & 17.0 & 14.86 &$-16.29$& 2   \\ 
    VCC\,1183 & IC 3413      & 12:29:22.49 & +11:26:01.8 & dS0,N  & 15.31 & 17.0 & 14.37 &$-16.78$&     \\ 
    VCC\,1261 & NGC 4482     & 12:30:10.35 & +10:46:46.3 & dE5,N  & 17.34 & 17.0 & 13.72 &$-17.43$& 1,2 \\ 
    VCC\,1431 & IC 3470      & 12:32:23.39 & +11:15:47.4 & dE0,N  &  8.14 & 17.0 & 14.38 &$-16.77$&     \\ 
    VCC\,1549 & IC 3510      & 12:34:14.85 & +11:04:18.1 & dE3,N  & 11.61 & 17.0 & 14.86 &$-16.29$&     \\ 
    VCC\,1695 & IC 3586      & 12:36:54.79 & +12:31:12.3 & dS0    & 13.72 & 17.0 & 14.67 &$-16.48$&     \\ 
    VCC\,1861 & IC 3652      & 12:40:58.60 & +11:11:04.1 & dE0,N  & 13.83 & 17.0 & 14.47 &$-16.68$&     \\ 
    VCC\,1910 & IC 809/3672  & 12:42:08.68 & +11:45:15.9 & dE1,N  &  ---  & 17.0 & 14.27 &$-16.88$&     \\ 
    VCC\,1912 & IC 810       & 12:42:09.12 & +12:35:48.8 & dS0,N  & 16.38 & 17.0 & 14.25 &$-16.90$&     \\ 
    VCC\,1947 & CGCG 043-003 & 12:42:56.36 & +03:40:35.6 & dE2,N  &  9.15 & 17.0 & 14.65 &$-16.50$& 1   \\ 
    \hline
    \hline
  \end{tabular}
\end{minipage}
\end{table*}
\begin{table}
  \caption{Observation log}
  \label{tab_obslog}
  \begin{tabular}{lrrl}
    \hline   
    galaxy     & night & P.A.  & exp \\
             &       & (deg) & (s) \\
    \hline
    M\,32     & 30 Dec 05 & 170 & $1 \times 600$ \\
    ID\,0650  &  4 Apr 06 & 140 & $4 \times 1200$ \\
    ID\,0734  & 29 Dec 05 &  15 & $5 \times 1200$ \\
    ID\,0872  & 30 Dec 05 & 105 & $9 \times 1200$ \\
    ID\,0918  &  4 Apr 06 & 175 & $4 \times 600$ \\
    ID\,1524  &  5 Apr 06 &  25 & $7 \times 1200$ \\
    VCC\,0021 &  6 Apr 06 & 100 & $4 \times 1200$ \\
    VCC\,0308 &  4 Apr 06 & 110 & $3 \times 1200$ \\
    VCC\,0397 &  5 Apr 06 & 135 & $4 \times 1200$ \\
    VCC\,0523 &  5 Apr 06 & 3$^a$ & $3 \times 1200$ \\
    VCC\,0856 & 29 Dec 05 & 105 & $3 \times 1200$ \\
    VCC\,0917 &  6 Apr 06 &  57 & $3 \times 1200$ \\
    VCC\,0990 &  4 Apr 06 & 135 & $3 \times 1200$ \\
    VCC\,1087 & 30 Dec 05 & 110 & $3 \times 1200$ \\
    VCC\,1122 &  4 Apr 06 & 132 & $3 \times 1200$ \\
    VCC\,1183 &  6 Apr 06 & 167 & $4 \times 1200$ \\
    VCC\,1261 & 29 Dec 05 & 145 & $3 \times 1200$ \\
    VCC\,1431 &  5 Apr 06 & 165 & $3 \times 1200$ \\
    VCC\,1549 &  5 Apr 06 &  10 & $3 \times 1200$ \\
    VCC\,1695 &  6 Apr 06 &  40 & $4 \times 1200$ \\
    VCC\,1861 & 30 Dec 05 &   0 & $5 \times 1200$ \\
    VCC\,1910 & 29 Dec 05 &   5 & $3 \times 1200$ \\
    VCC\,1912 &  4 Apr 06 & 166 & $3 \times 1200$ \\
    VCC\,1947 &  6 Apr 06 & 125 & $3 \times 1200$ \\
    \hline
    \hline
  \end{tabular}\\
  $^a$ The major axis position angle of VCC\,523 is 175\,deg. 
  We chose 3\,deg to also include VCC\,522 in the slit.
\end{table}
\begin{figure}
  \includegraphics[clip,width=8cm]{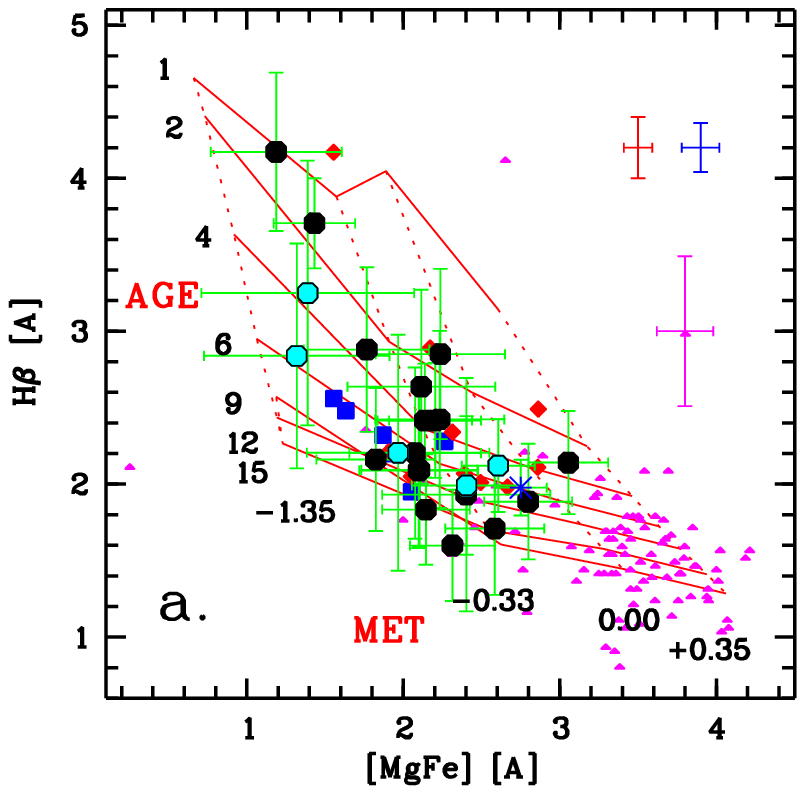}\\
  \includegraphics[clip,width=8cm]{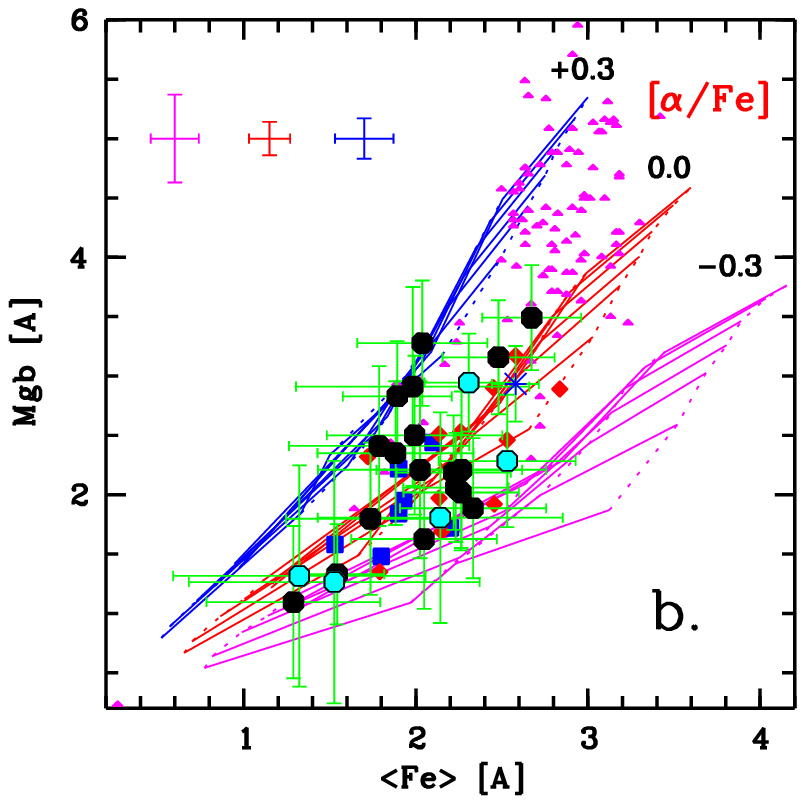}
  \caption{%
    Comparison of our dEs (Virgo: black filled circles, field: cyan filled
    circles, M\,32: blue asterisk) with the dE samples of \citet{gehaetal03}
    (red diamonds) and \citet{vanzeeetal04} (blue squares), and with the
    massive early-type galaxies from SB06 (magenta triangles). The average
    error bars for these other samples are plotted in the same color scheme.
    Overlaid are TMB03 models with different ages, metallicities and abundance
    ratios, as indicated on the figure. \textit{a.}~\Hbeta\ -- [MgFe] diagram
    overlaid with [$\alpha$/Fe]$ = 0$ model grid. \textit{b.}~Mg$b$ -- \Fe\ 
    diagram with different [$\alpha$/Fe] models.}
  \label{fig_Hbeta_MgFe_our_geha_vzee}
\end{figure}

\subsection{Sample}

A detailed description of the sample selection for the MAGPOP-ITP dwarf
galaxies will be presented in \citet{peletier07}. In summary, the galaxies
were selected to have been observed, but not necessarily detected, by GALEX.
For the Virgo sample we then selected galaxies with $m_B > 15$, classified as
dE or dS0 in the Virgo Cluster Catalogue \citep[VCC:][]{binggelietal85}. This
yields 43 objects, of which we observed 18, giving preference to those with
highest central surface brightness.  Therefore, all but two of the observed
dEs are nucleated (dE,N; see Table~\ref{tab_sample}), i.e. they have an
unresolved central light excess \citep{binggelietal85}.

For the field sample we queried SDSS for nearby dwarf galaxies ($0.00125 < z <
0.00625$ and $-18.5 < M_{r^\prime} < -15$\,mag)\footnote{The absolute
  magnitudes were computed using the SDSS radial velocities and assuming a
  Hubble constant $H_0 = 70\,\kms\,\textrm{Mpc}^{-1}$.}. To select quiescent
dwarf galaxies, we then applied a colour cut in UV colours (GALEX:
$\textrm{FUV} - \textrm{NUV} > 0.9$), or in optical colours (SDSS: $u - g >
1.2$) if there where non-detections in the UV. These colour cuts maximise the
separation in star-forming and quiescent galaxies in the Virgo sample
\citep[see][]{peletier07}. Visual inspection of this selected sample yielded
10 objects.  However, from the SDSS spectroscopic data, we found that three of
these 10 contain emission lines.  Because emission lines are hard to remove
accurately from the intermediate-resolution spectra we analyse here, we did
not include these galaxies in the sample presented here. One of them
(NGC~3073) will be the subject of an extensive analysis based on
high-resolution spectroscopy \citep{toloba07}. We were mostly limited by
visibility constraints and observed only 5 field dEs.

Finally, we also observed M~32 to compare to previous studies. In
Table~\ref{tab_sample}, the sample galaxies and some of their properties are
listed. In column~1 we give the name of the galaxies, either by their number
in the VCC catalogue, or by their GALEX identification number (ID), while in
column 2 we give alternative names. Columns~3 and 4 list the galaxy positions.
For column~5 we take the morphological type given by NED (for the field dEs)
or by VCC (for the Virgo dEs). Column~6 gives, if available, the $H$-band
effective radius in arcseconds, taken from the GOLDMine database
\citep{gavazzi03goldmine}.  Columns~7 lists the distance $D$ to the galaxies.
For M~32 we take the distance from the HYPERLEDA database.
\citep{paturel03hyperleda}.  For the field galaxies we calculate the distance
based on their radial velocity and assuming $H_0 =
70\,\kms\,\textrm{Mpc}^{-1}$. The distances to the Virgo galaxies are
estimated using their position in the Virgo Cluster \citep{gavazzietal99}. The
apparent magnutides $B_T$ (column~8) are taken from the HYPERLEDA database,
and the absolute blue magnitudes $M_B$ (column~9) are computed using the
listed distances and apparent magnitudes. Finally, in column~10 we indicate
which galaxies have also been observed by \citet{gehaetal03} (1), and by
\citet{vanzeeetal04} (2).

\subsection{Observations and data reduction}

The observations were carried out on 29--30 December 2005 and 4--6 April 2006
with the 2.5m Nordic Optical Telescope (NOT) using ALFOSC with grism \#14
(600\,rules\,mm$^{-1}$ and blazed at $\lambda=4288$\,\AA) and
slit\footnote{Although the name of the used slit is 1.2\arcsec, the actual
  measured width is 1.0\arcsec. See the ALFOSC page for more details:
  http://www.not.iac.es/instruments/alfosc/slits.html} 1.2\arcsec. The
wavelength coverage is $\lambda\lambda 3240 - 6090$\,\AA\ and the resolution
is 6.8\,\AA\ (FWHM), or $\sigma_{\rm instr} = 170$\,\kms\ around 5200\,\AA.
The detector was an E2V Technologies back-illuminated CCD with
$2048\times2048$ active + 50 overscan pixels on both sides, with a pixel size
of $13.5\,\mu$m, and a plate scale of $0.19\arcsec$\,pixel$^{-1}$. At the
beginning of most of the nights there was thin cirrus, and the seeing varied
between 0.8--1.5\arcsec (FWHM). Typical integration times were from one hour,
up to 3 hours for the faintest galaxies (see Table~\ref{tab_obslog}).

The data reduction was carried out using MIDAS\footnote{The image processing
  package ESO-MIDAS is developed and maintained by the European Southern
  Observatory.} and
REDUCEME\footnote{http://www.ucm.es/info/Astrof/software/reduceme/reduceme.html}
\citep{cardiel99}. All frames were overscan and bias subtracted, flat-fielded
using dome and twilight flats and cosmic ray events were removed. For the
wavelength calibration we obtained He arcs at each telescope position.

During twilight we observed standard stars drawn from the Lick/IDS
\citep{wortheyetal94} and MILES \citep{sanblas06miles}stellar libraries,
spanning a range of spectral types (see Table~\ref{tab_standardstars}). Since
the MILES catalogue is carefully flux-calibrated, we used these stars as
spectrophotometric standards, as well as velocity templates. Finally we
de-redshifted the galaxy spectra.
Given the spectral resolution of $\sigma_{\rm instr} \sim 170$\,\kms\ and the
expected galactic velocity dispersions, $\sigma_{\rm gal} \sim 50$\,\kms, it
is not possible to measure the internal dispersions. A kinematic analysis is
therefore beyond the scope of this paper and will be addressed in a subsequent
paper \citep{toloba07} using high-resolution spectroscopy.  All galaxies have
been observed with the slit along the major axis, determined from SDSS surface
photometry analysis, except for VCC\,0523, where we positioned the slit to
simultaneously observe VCC\,0522 (see Table~\ref{tab_obslog}).

\begin{figure*}
  \includegraphics[clip,width=0.95\textwidth]{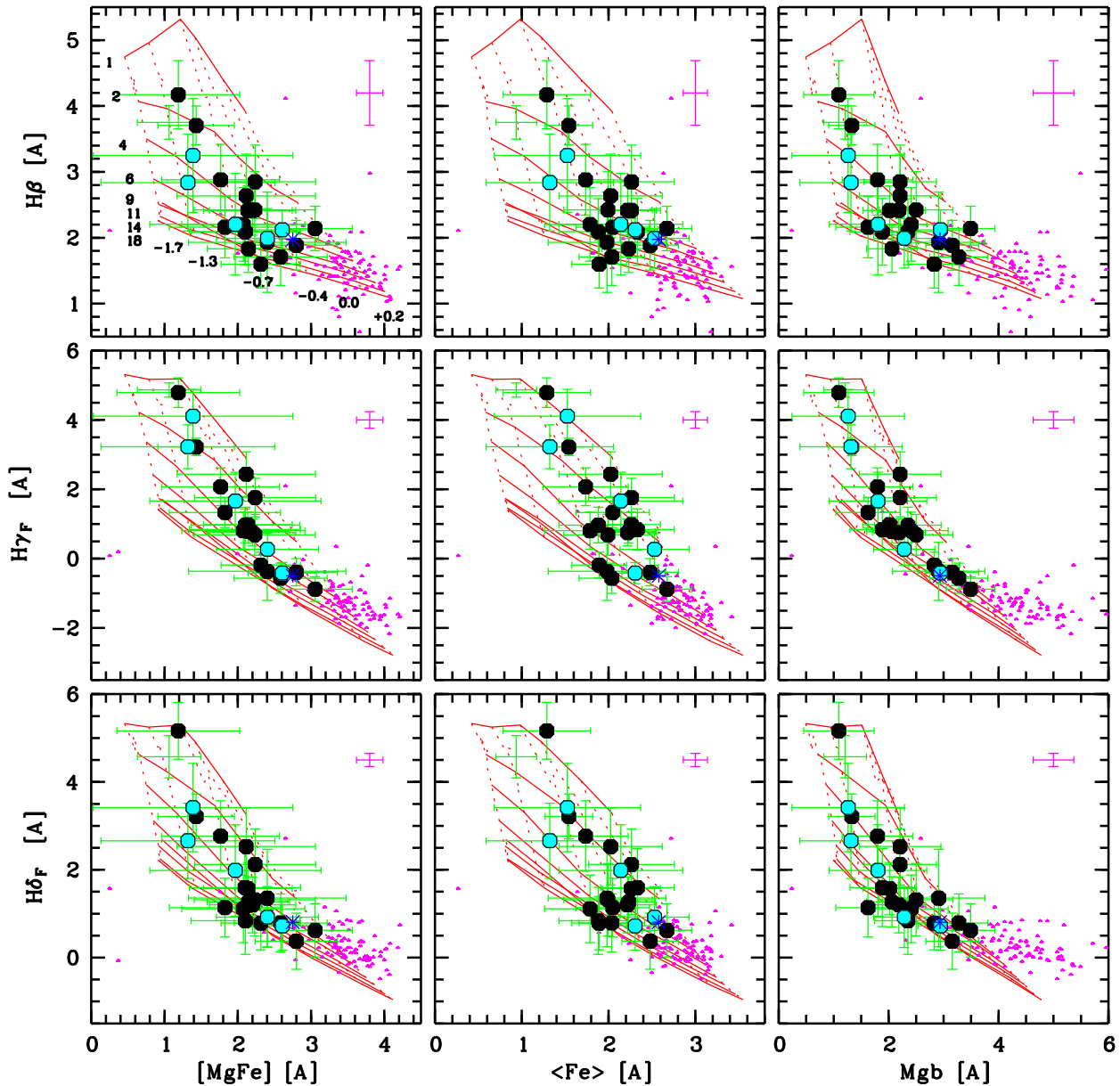}
  \caption{%
    Lick/IDS index-index diagrams of age-sensitive Balmer \Hbeta, \HgF\ and
    \HdF\ versus metallicity-sensitive Mg$b$, \Fe\ and [MgFe] for our dEs and
    SB06 normal Es (the mean error for this data is shown as a cross in the
    bottom-left corner of each panel). The symbols are the same as in
    Figure~\ref{fig_Hbeta_MgFe_our_geha_vzee}. The model grids are those of
    V96.}
  \label{fig_Balmer_MgFe}
\end{figure*}

\subsection{Index calibration}

In this paper, we work with flux-calibrated, one-dimensional spectra, obtained
by summing the central 4\arcsec\ in each galaxy. Given the typical effective
radius $r_{\rm eff} \approx 8 - 20\arcsec$, this corresponds to about $r_{\rm
  eff}/4 - r_{\rm eff}/10$. To be able to work with the information contained
in the spectra, we have determined Lick/IDS indices, allowing us easy
comparisons with the literature and some stellar population models (see
below).  However, the reduced spectra are also available on simple request,
allowing the reader to measure their own indices at their preferred spectral
resolution, or to use all the available information.

To measure indices in the Lick/IDS system, we broadened our spectra to the
Lick/IDS resolution ($\sim 8.4-10$\,\AA\ (FWHM), depending on the wavelength,
see Table~\ref{tab_offsets}).  Then we measure indices using the pass-bands
defined in \citet{wortheyetal94} and \citet{wortheyottaviani97}. We observed
standard stars from in the original Lick/IDS stellar library
(Table~\ref{tab_standardstars}) to determine possible systematic offsets
resulting from the non-flux-calibrated response of the Lick/IDS system
(Table~\ref{tab_offsets}). See Appendix~\ref{app_transformation} for more
details on the transformation to the Lick/IDS system and
Appendix~\ref{app_indices} for a table with all the measured indices. In none
of the galaxies in the sample do we detect [OIII] emission, thus we do not
correct the \Hbeta\ absorption for possible contamination by emission. If such
emission were present, it would make the measured \Hbeta\ absorption smaller,
and therefore the derived ages older.

To derive ages, metallicities and abundance ratios, we use predictions
of single-age, single-metallicity stellar population (or simple
stellar population or SSP) models, in particular those of
\citet{TMB03} (TMB03) and \citet{vazdekis96} (V96), as updated in
\citet{vazdekis99} and \citet{vazdekis03}. One should keep in mind
that SSPs that are based on observational stellar libraries might not
be solar-scaled at every metallicity, because of the limitations of
the input library. One could expect that models are solar-scaled at
high metallicity because here stars in the solar neighbourhood are
being used, while at low metallicity the models may be
$\alpha$-enhanced because low-metallicity stars in the observational
library are generally somewhat $\alpha$-enhanced \citep{maraston03}.
TMB03 tried to correct for this bias by assuming an $\alpha$/Fe -
metallicity relation for the abundances of the input stars. The
underlying stellar isochrones in both the V96 and TMB03 models are
scaled solar for all metallicities. It is instructive to see how well
the results obtained using different models agree, and that the choice
of model does not bias the conclusions obtained here in any way.

\subsection{Comparison data}

In order to compare the behaviour of dEs with that of normal Es, we
use the sample of 98 early-type galaxies from \citet{sanblas06}
(SB06), who provided us with indices measured within the central
4\arcsec\ \citep{sanblas04thesis}.  This sample contains a range of
early-type galaxies of all luminosities in dense (Coma) and less dense
(Virgo and field) environments.

We also examined the dE samples of \citet{vanzeeetal04} and
\citet{gehaetal03}. The \citet{vanzeeetal04} sample comprises 16 dE/dS0 from
the VCC with $m_B < 15.5$.  Because they were looking for rotationally
supported dEs, an ellipticity constraint $\epsilon > 0.25$ was also imposed.
The sample of \citet{gehaetal03} is also selected from the bright end of the
dE population in the Virgo Cluster, spanning a range of ellipticities, and
consists of 17 objects. The three dE samples span a similar range of
magnitudes and there is an overlap of 4--5 galaxies between each of the
samples. See Table~\ref{tab_sample} for dEs in common with our sample, and
Appendix~\ref{app_transformation} (Figure~\ref{fig_compare_our_geha_vzee}) for
a comparison of the indices measured in the three papers. Both
\citet{gehaetal03} and \citet{vanzeeetal04} find no statistical differences
between the rotating and the non-rotating dEs.  Although we will use here some
preliminary results from \citep{toloba07} to identify dEs with rotation in our
sample (see Section~\ref{sec_virgodistance}), we postpone a detailed
comparison of the rotating and non-rotating dEs to that paper.

\section{Results}
\label{sec_results}

The ultimate goal of our analysis is to gain insight in the star formation and
assembly history of dwarf galaxies. This can be done by comparing indices, and
combinations of indices, that are sensitive to age, metallicity or relative
abundance of different metals. The measured values are compared to population
synthesis model predictions. Comparison with SSP models gives us
SSP-equivalent, or \emph{mean, luminosity-weighted}, ages, metallicities and
abundance ratios. In the optical most of the light comes from the youngest
component of the stellar population. It is important to keep this in mind in
the following. A galaxy with a \emph{young age} means that the galaxy formed
stars until recently, but could have a very old underlying stellar population.
Throughout the paper, when we talk about age, metallicity or abundance ratio,
we refer to the mean, luminosity-weighted values.

\subsection{Central Lick/IDS indices}
\label{sec_indices}
\begin{figure}
  \includegraphics[clip,width=8cm]{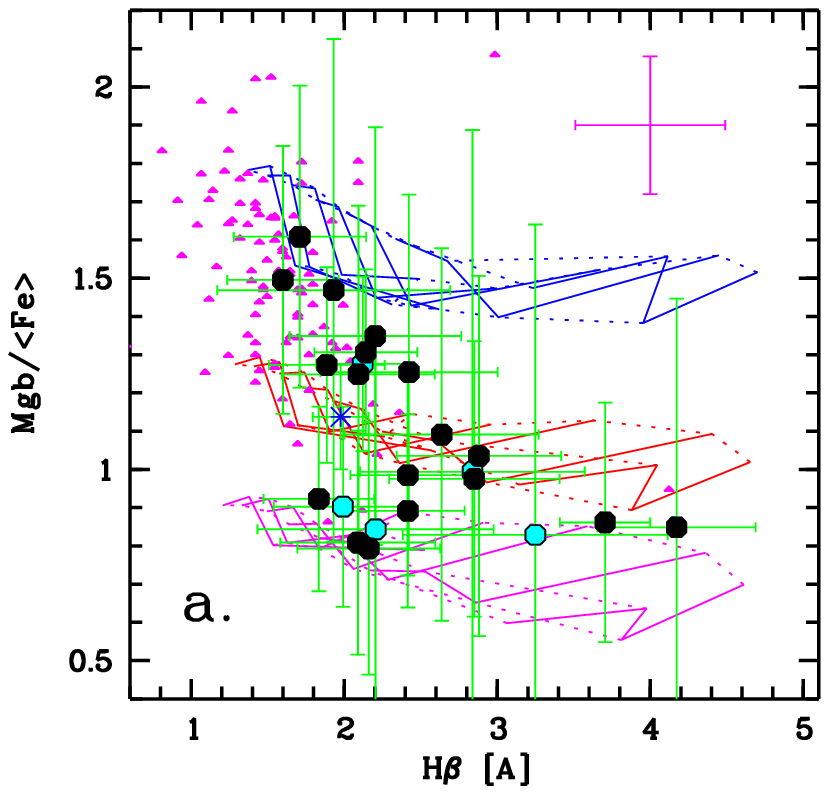}\\
  \includegraphics[clip,width=8cm]{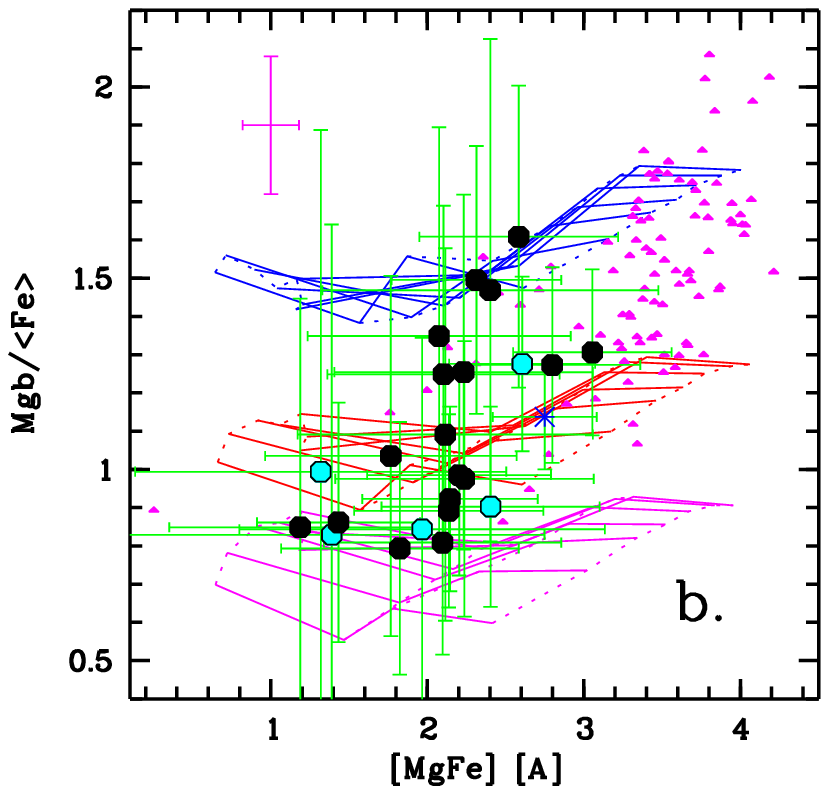}
  \caption{%
    The index ratio Mg$b$/\Fe\ is mostly sensitive to the [$\alpha$/Fe]
    abundance ratio. \textit{a.}~Mg$b$/\Fe\ versus \Hbeta\ and
    \textit{b.}~Mg$b$/\Fe\ versus [MgFe] for our dEs and SB06 Es, overlaid
    with TMB03 models (same symbols and models as in
    Figure~\ref{fig_Hbeta_MgFe_our_geha_vzee}).}
  \label{fig_MgoverFe_our_sanblas}
\end{figure}
\begin{figure*}
  \includegraphics[clip,width=16cm]{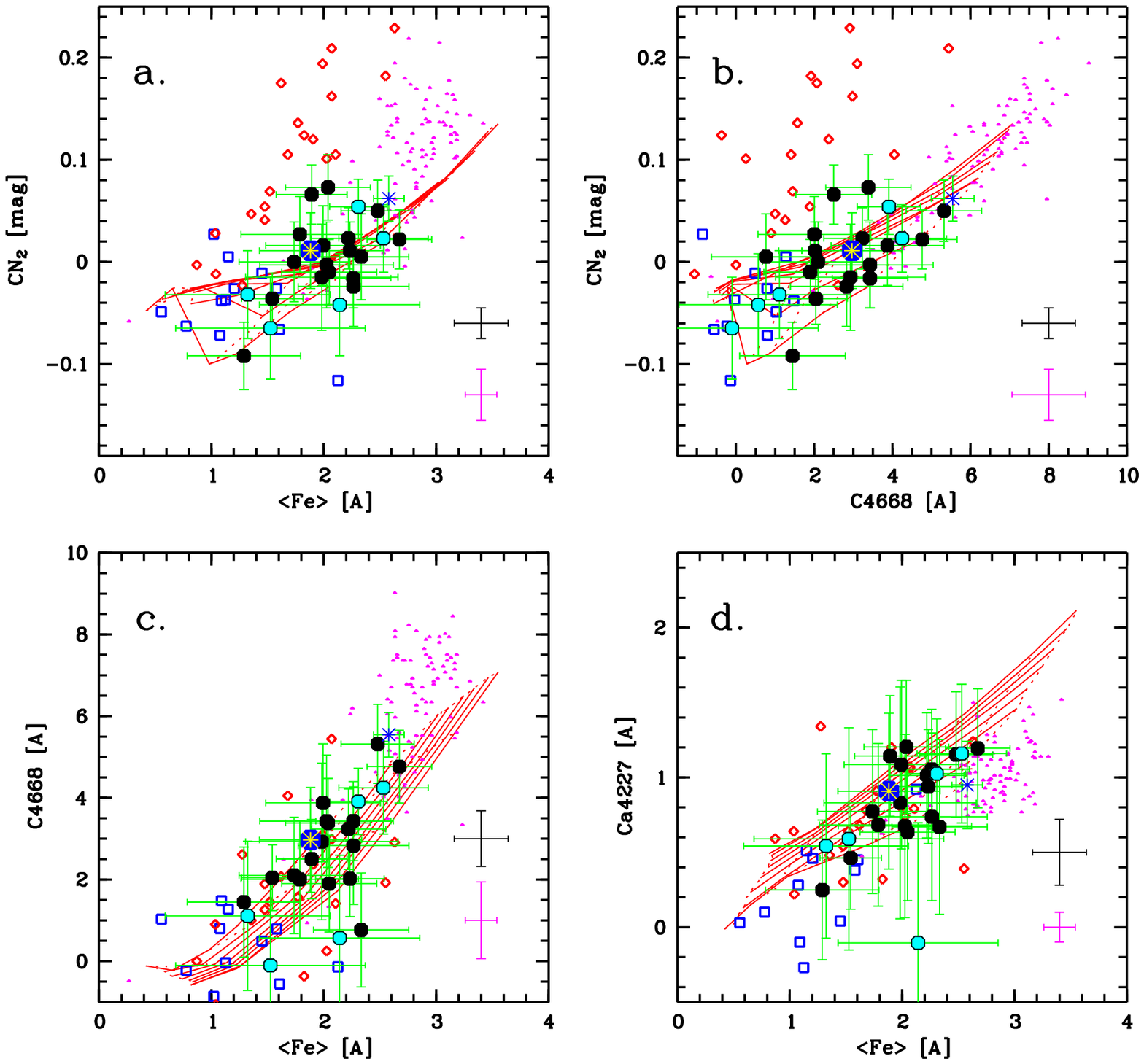}
  \caption{%
    Comparison of CN$_2$, C4668 and Ca4227 versus \Fe. Superimposed are V96
    SSP models . The symbols are the same as in
    Figure~\ref{fig_Hbeta_MgFe_our_geha_vzee}. We also plot the globular
    cluster systems of the giant E NGC~1407 \citep[][red open
    diamonds]{cenarroetal07} and the dE VCC\,1087 \citep[][blue open
    squares]{beasleyetal06}. VCC\,1087 is in our sample and is highlighted in
    blue overplotted with a yellow asterisk. At the bottom right of each
    panel, the magenta error bars show the mean errors for the SB06 sample,
    whereas the black error bars show the mean errors for the globular cluster
    data.}
  \label{fig_Fe_metals}
\end{figure*}
\begin{figure*}
  \includegraphics[clip,width=16cm]{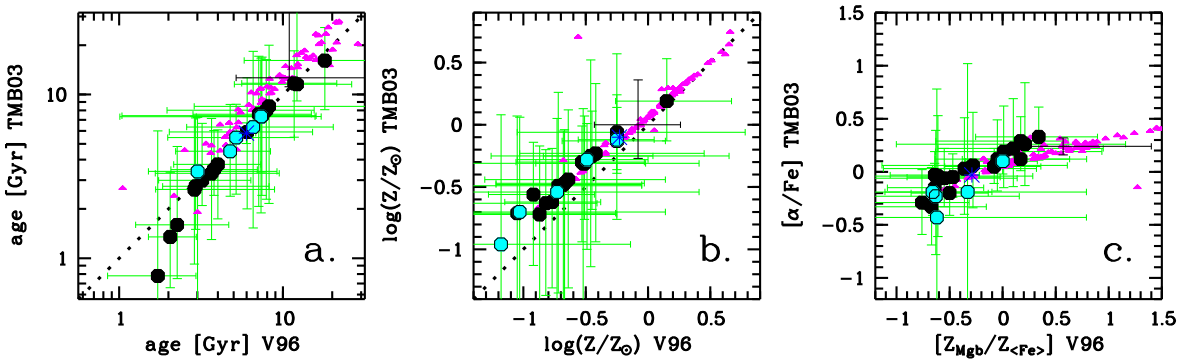}
  \caption{%
    Comparison of age, metallicity and [$\alpha$/Fe] abundance ratio derived
    using V96 models and TMB03 models. Symbols are the same as in
    Figure~\ref{fig_Hbeta_MgFe_our_geha_vzee}. The black error bars show the
    errors for a typical galaxy in the SB06 sample.}
  \label{fig_compare_vaz_TMB}
\end{figure*}
\begin{figure*}
  \includegraphics[clip,width=16cm]{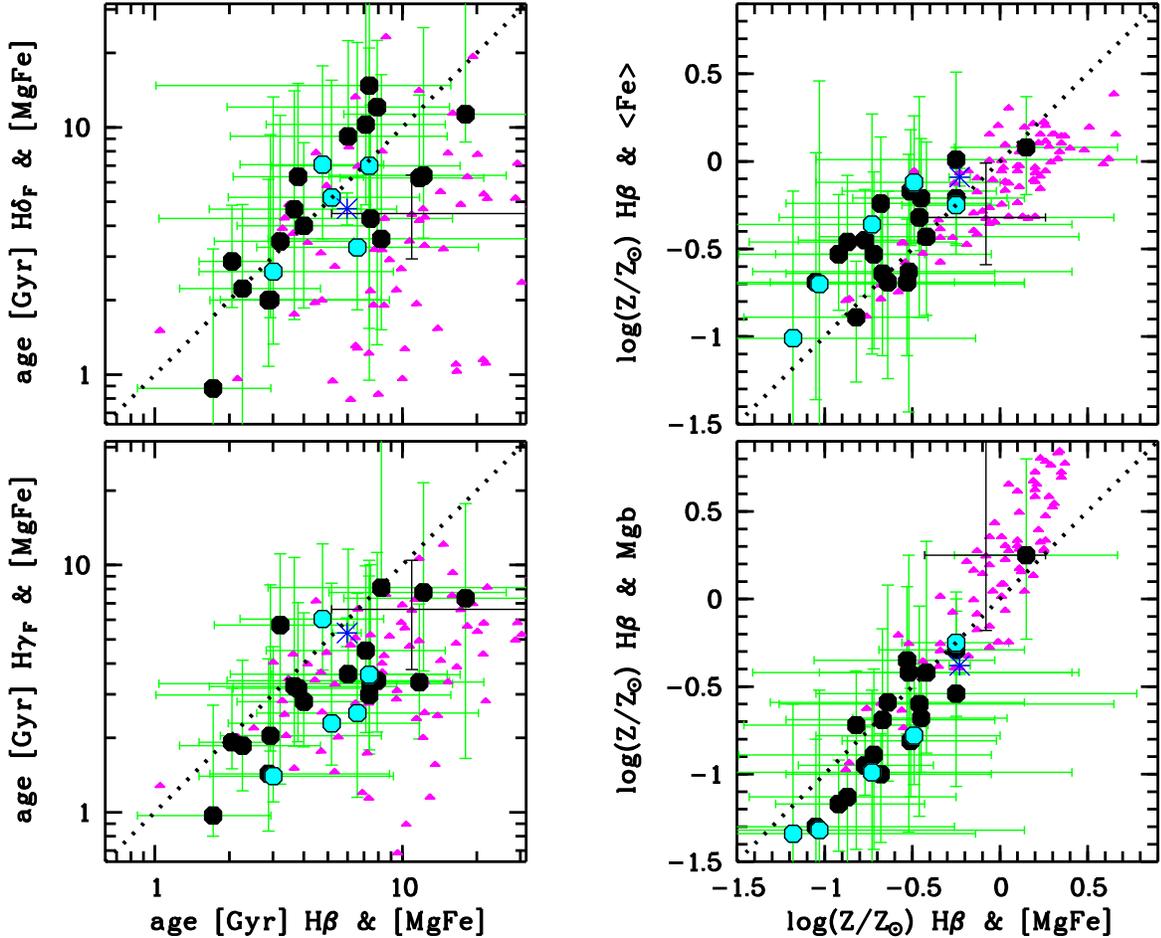}
  \caption{%
    Comparison of ages and metallicities derived from different age and
    metallicity indicators, and using V96 models.  \textit{Left:} Ages from
    \Hbeta, \HgF\ and \HdF\ versus [MgFe] diagrams.  \textit{Right:}
    Metallicities from \Hbeta\ versus [MgFe], Mg$b$ and \Fe\ diagrams. Symbols
    are the same as in Figure~\ref{fig_Hbeta_MgFe_our_geha_vzee}. The black
    error bars are typical error bars on a galaxy in the SB06 sample. The
    dotted line is the 1--1 relation.}
  \label{fig_compare_agemet}
\end{figure*}

The most age-sensitive Lick/IDS indices are the hydrogen Balmer series, of
which \Hbeta, H$\gamma$ and H$\delta$ are measurable in our spectra. The most
used metallicity-sensitive Lick/IDS indices are Mg$b$, Fe5270 and Fe5335,
often combined as $\textrm{\Fe} = (\textrm{Fe5270} + \textrm{Fe5335}) / 2$,
and $\textrm{[MgFe]} = \sqrt{ \textrm{Mg$b$} \times \textrm{\Fe} }$. The Mg$b$
index traces the metallicity as given by the $\alpha$ elements, whereas \Fe\ 
is most sensitive to Fe. Discrepancies between the metallicity estimated by
those indices are due to non-solar [$\alpha$/Fe] abundance ratios
\citep{wortheyetal92}. The [MgFe] index tries to minimise the effect of the
non-solar abundance ratios (especially Mg-enhancement) exhibited by normal Es
(TMB03).

In Figure~\ref{fig_Hbeta_MgFe_our_geha_vzee} we plot the \Hbeta -- [MgFe] and
Mg$b$ -- \Fe\ diagrams for our sample of dEs, together with the dEs of
\citet{gehaetal03} and \citet{vanzeeetal04}, and the early-type massive
galaxies of SB06. The dE data fall in the same region in these plots,
indicating that they have similar stellar population properties---ages,
metallicities and abundance ratios. Comparing to the TMB03 models,
Figure~\ref{fig_Hbeta_MgFe_our_geha_vzee}a shows that dEs span a wide range of
ages and metallicities. Note however, that there appear to be no old,
metal-rich dEs. The dEs also have more or less solar abundance ratios and some
have even sub-solar abundance ratios
(Figure~\ref{fig_Hbeta_MgFe_our_geha_vzee}b), a fact that was already noted by
\citet{vanzeeetal04} and confirmed here. In the following we leave out the
other dE samples because they would make the plots too crowded, and we
concentrate on the comparison between dEs and Es.

Figure~\ref{fig_Hbeta_MgFe_our_geha_vzee}a shows that the massive Es
completely fill the bottom-right part of the \Hbeta\ -- [MgFe] model grid,
demonstrating that they are mainly old, metal-rich systems. In the Mg$b$ --
\Fe\ diagram (Figure~\ref{fig_Hbeta_MgFe_our_geha_vzee}b) the dEs and Es form
a continuum of increasingly higher abundance ratios.

In Figure~\ref{fig_Balmer_MgFe} we compare the observed central
\Hbeta, H$\gamma$ and H$\delta$ indices versus Mg$b$, \Fe\ and [MgFe]
indices to the V96 models. Again, while the massive early-type
galaxies fill the bottom-right corner of the model grids, the dEs show
a much larger spread, practically filling the rest of the grid,
although the error bars are fairly large. There is almost no overlap
between the dEs and the Es, but they seem to form a continuum. The
Mg$b$ values measured in normal Es fall outside the grids because of
the super-solar [$\alpha$/Fe] abundance ratios in these massive
systems. The dEs however, fall in the same region as the model grids
in both \Fe\ and Mg$b$ and any of the Balmer indices. Finally it
should be noted that the normal Es also have different loci with
respect to the model grids when looking at the Balmer indices. This
reflects the influence of super-solar [$\alpha$/Fe] abundance ratios
on the higher-order Balmer indices \citep[see][]{thomasetal04}. The
\Hbeta\ and [MgFe] indices are virtually independent on the
[$\alpha$/Fe] abundance ratio.  Therefore we will use these indices to
derive the ages and metallicities of the galaxies (unless otherwise
specified, see section~\ref{sec_agemetalfa}).

The Mg$b$ and the \Fe\ index are sensitive to both metallicity and
[$\alpha$/Fe] abundance ratio. The combination into [MgFe] makes a
good metallicity estimator that is virtually independent of the
[$\alpha$/Fe] ratio (TMB03). The ratio Mg$b$/\Fe\ on the other hand is
almost independent of the metallicity. In
Figure~\ref{fig_MgoverFe_our_sanblas} we plot Mg$b$/\Fe\ versus
\Hbeta\ and [MgFe]. Unfortunately, error bars tend to explode when
computing line-strength ratios, but as in
Figure~\ref{fig_Hbeta_MgFe_our_geha_vzee}b, the Es clearly fall above
any solar-scaled models, whereas the dE abundance ratios scatter
around solar; some even show sub-solar ratios. Again, there appears to
be a continuum from Es to dEs: decreasing Mg$b$/\Fe\ (decreasing
[$\alpha$/Fe] abundance ratio) with increasing \Hbeta\ (decreasing
age), and with decreasing [MgFe] (decreasing metallicity). It seems
that, whilst for the oldest dEs there is a range of [$\alpha$/Fe]
ratios, all the youngest dEs (${\rm H}\beta > 2.5$\,\AA) have solar or
sub-solar abundance ratios.

\subsubsection{Other metal indicators}

Since different elements are produced on different timescales, element
abundance ratios can potentially be used as clocks that measure the
duration of the star formation in a galaxy. By comparing other metal
indices to \Fe, we can gain insight in the different abundance ratios
on a relative scale. We do not attempt to obtain the true abundance
ratios as these are always model-dependent. It is our opinion that
current-day stellar population models are not capable of deriving true
abundance ratios.

In Figure~\ref{fig_Fe_metals} we show CN$_2$ versus \Fe\ and C4668
(panels~a and b respectively) as well as C4668 and Ca4227 versus \Fe\
(panels~c and d). The models plotted in Figure~\ref{fig_Fe_metals} are
those of V96. We also checked with the solar-scaled TMB03 models and
found that not correcting the $\alpha$-enhancement (V96) has some
effect at very low metallicities but it is not significant and does
not change any of our results.

In \citet{tripiccobell95}, it was established that the C4668 and
Ca4227 indices are mostly sensitive to C and Ca respectively. For the
CN$_2$ index, the above paper predicts a C and N dependence which
varies with stellar type. However, the observation that CN and NH
features in M31 globular clusters (GCs) are enhanced with respect to
Milky Way GCs, whilst the CH feature is not, supports the idea that N
rather than C, is actually driving the CN$_2$ values
\citet[][see]{bursteinetal84, bursteinetal04}.

The behaviour of N and C in the integrated spectra of stellar
populations has turned out to be a promising tool to constrain not
only different star formation time-scales but also the importance of
different chemical enrichment processes \citep[see][and references
therein]{cenarroetal07}. For a more general comparing overview with
other subsamples, in Figure~\ref{fig_Fe_metals} we included data for
the globular cluster systems of the giant E NGC~1407
\citep{cenarroetal07} and the dE VCC\,1087 \citep[with the galaxy
being also in our sample]{beasleyetal06}. Both globular cluster data
sets have been spectroscopically confirmed to be old ($> 10$\,Gyr).
Interestingly, VCC\,1087 is slightly younger with an age of 7.4\,Gyr
\citep[in agreement with][]{beasleyetal06}, as will be presented in
next Section.

In panels~a and b, most dEs exhibit striking N underabundances with respect to
Fe and C, contrary to both massive Es and extragalactic globular clusters of
similar metallicity. Note however that, whilst dEs seem to match the
extrapolation of the metallicity sequence of massive Es down to lower values,
globular clusters clearly depart from the locus of galaxies all over the
metallicity regime. Probably, primordial N enhancements in globular clusters
\citep{meynetmaeder02, liburstein03} are responsible for this dichotomy. In
panel~c, the dichotomy between galaxies and globular cluster has washed out.
Instead, from massive Es down to the low metallicity globular clusters there
exists a unique sequence in the sense that [C/Fe] ratios tend to increase with
the increasing metallicity. Note however that, since age differences among the
subsamples exist, this trend must be considered from a qualitative point of
view. In any case, it seems clear that most dEs show slightly super-solar
[C/Fe] ratios---resembling massive Es---whereas globular clusters tend to
exhibit either solar or even subsolar values \citep[see][]{cenarroetal07}.

Ca has received recent attention in the literature because of inconsistencies
found for the predictions of the NIR Ca II Triplet
\citep[e.g.][]{cenarroetal03, michielsenetal03}.  Also the optical Ca4227
index exhibits unexplained behaviour \citep{cenarroetal04}. Although Ca is an
$\alpha$ element, we see that in Figure~\ref{fig_Fe_metals}d the indices for
both dEs and Es are consistent with solar [Ca/Fe] abundance ratios. This could
point to a genuine Ca depletion resulting from metallicity-dependent supernova
yields \citep{thomasetal03}.  However, recently \citet{prochaskaetal05} have
indicated that the CN band might be affecting the blue pseudo-continuum of the
Ca line, leading to lower Ca4227 values.  Therefore, giant Es, which show
strong CN should also show low Ca4227 values, as observed here. The dEs on the
other hand have no strong CN absorption. We therefore do not expect a
CN-induced decrease in the Ca4227 index. The observed trend that dEs have
solar [Ca/Fe] ratios is therefore real. In a subsequent paper we will
investigate these issues in more detail, using high-resolution spectra and SSP
models \citep{toloba07}.

\subsection{Ages, metallicities and abundance ratios}
\label{sec_agemetalfa}

\begin{table*}
  \caption{Ages and metallicities from different indices using V96 models} 
  \label{tab_agemet_V96}
  \begin{tabular}{l*6{r@{}l}}
    \hline
             & \multicolumn{4}{c}{\Hbeta\ -- [MgFe]} & \multicolumn{4}{c}{\Hbeta\ -- Mg$b$} & \multicolumn{4}{c}{\Hbeta\ -- \Fe} \\
    galaxy     & \multicolumn{2}{c}{age} & \multicolumn{2}{c}{log(Z/Z$_\odot$)} & \multicolumn{2}{c}{age} & \multicolumn{2}{c}{log(Z/Z$_\odot$)} & \multicolumn{2}{c}{age} & \multicolumn{2}{c}{log(Z/Z$_\odot$)} \\
             & \multicolumn{2}{c}{(Gyr)} & \multicolumn{2}{c}{(dex)} & \multicolumn{2}{c}{(Gyr)} & \multicolumn{2}{c}{(dex)} & \multicolumn{2}{c}{(Gyr)} & \multicolumn{2}{c}{(dex)} \\
    \hline
    M\,32     & $ 5.98$ & $^{+ 0.84}_{- 0.76}$ & $-0.23$ & $^{+0.06}_{-0.06}$ & $ 6.87$ & $^{+ 0.89}_{- 0.80}$ & $-0.38$ & $^{+0.05}_{-0.05}$ & $ 5.20$ & $^{+ 0.70}_{- 0.64}$ & $-0.09$ & $^{+0.05}_{-0.05}$ \\
    ID\,0650  & $ 7.35$ & $^{+ 9.78}_{- 4.39}$ & $-0.49$ & $^{+0.49}_{-0.56}$ & $ 9.22$ & $^{+ 7.11}_{- 5.90}$ & $-0.78$ & $^{+0.41}_{-0.28}$ & $ 5.26$ & $^{+ 7.27}_{- 3.46}$ & $-0.12$ & $^{+0.38}_{-0.35}$ \\
    ID\,0734  & $ 6.58$ & $^{+13.72}_{- 4.00}$ & $-0.73$ & $^{+1.14}_{-0.95}$ & $ 8.35$ & $^{+19.93}_{- 4.44}$ & $-0.99$ & $^{+0.47}_{-0.79}$ & $ 4.61$ & $^{+12.87}_{- 3.75}$ & $-0.36$ & $^{+0.63}_{-0.74}$ \\
    ID\,0872  & $ 3.01$ & $^{+ 6.19}_{- 1.50}$ & $-1.03$ & $^{+1.17}_{-1.50}$ & $ 4.03$ & $^{+ 5.94}_{- 2.36}$ & $-1.32$ & $^{+0.80}_{-1.05}$ & $ 2.39$ & $^{+ 4.14}_{- 1.09}$ & $-0.70$ & $^{+1.16}_{-0.94}$ \\
    ID\,0918  & $ 4.76$ & $^{+ 3.64}_{- 2.55}$ & $-0.25$ & $^{+0.30}_{-0.31}$ & $ 4.79$ & $^{+ 3.67}_{- 2.58}$ & $-0.25$ & $^{+0.25}_{-0.27}$ & $ 4.74$ & $^{+ 3.32}_{- 2.22}$ & $-0.25$ & $^{+0.21}_{-0.22}$ \\
    ID\,1524  & $ 5.18$ & $^{+ 6.56}_{- 3.20}$ & $-1.18$ & $^{+1.04}_{-1.29}$ & $ 5.70$ & $^{+ 6.15}_{- 4.24}$ & $-1.34$ & $^{+0.74}_{-0.96}$ & $ 4.49$ & $^{+ 6.25}_{- 3.53}$ & $-1.01$ & $^{+0.84}_{-0.84}$ \\
    VCC\,0021 & $ 1.72$ & $^{+ 1.22}_{- 0.87}$ & $-1.05$ & $^{+0.79}_{-0.83}$ & $ 1.90$ & $^{+ 1.13}_{- 0.69}$ & $-1.30$ & $^{+0.50}_{-0.53}$ & $ 1.31$ & $^{+ 1.12}_{- 0.28}$ & $-0.69$ & $^{+0.74}_{-0.67}$ \\
    VCC\,0308 & $ 2.88$ & $^{+ 6.00}_{- 1.21}$ & $-0.46$ & $^{+1.11}_{-0.80}$ & $ 3.42$ & $^{+ 5.26}_{- 1.46}$ & $-0.60$ & $^{+0.56}_{-0.64}$ & $ 2.53$ & $^{+ 4.79}_{- 0.79}$ & $-0.32$ & $^{+0.69}_{-0.57}$ \\
    VCC\,0397 & $ 2.26$ & $^{+ 2.41}_{- 1.00}$ & $-0.25$ & $^{+1.03}_{-0.80}$ & $ 2.64$ & $^{+ 3.32}_{- 0.77}$ & $-0.54$ & $^{+0.47}_{-0.53}$ & $ 1.96$ & $^{+ 1.42}_{- 0.71}$ & $ 0.01$ & $^{+0.50}_{-0.54}$ \\
    VCC\,0523 & $ 3.66$ & $^{+ 5.82}_{- 2.00}$ & $-0.42$ & $^{+0.87}_{-0.61}$ & $ 3.68$ & $^{+ 4.74}_{- 2.22}$ & $-0.42$ & $^{+0.75}_{-0.46}$ & $ 3.65$ & $^{+ 5.37}_{- 1.89}$ & $-0.43$ & $^{+0.54}_{-0.45}$ \\
    VCC\,0856 & $ 6.04$ & $^{+ 9.12}_{- 4.02}$ & $-0.64$ & $^{+0.78}_{-0.67}$ & $ 5.74$ & $^{+ 8.48}_{- 3.69}$ & $-0.59$ & $^{+0.67}_{-0.45}$ & $ 6.33$ & $^{+ 9.17}_{- 5.30}$ & $-0.69$ & $^{+0.52}_{-0.55}$ \\
    VCC\,0917 & $ 7.44$ & $^{+ 8.49}_{- 6.40}$ & $-0.68$ & $^{+0.63}_{-0.54}$ & $13.11$ & $^{+29.86}_{- 1.14}$ & $-1.00$ & $^{+0.23}_{-0.39}$ & $ 4.97$ & $^{+ 7.49}_{- 3.53}$ & $-0.24$ & $^{+0.38}_{-0.43}$ \\
    VCC\,0990 & $11.71$ & $^{+ 9.62}_{- 6.11}$ & $-0.77$ & $^{+0.39}_{-0.38}$ & $12.54$ & $^{+ 5.53}_{- 4.80}$ & $-0.95$ & $^{+0.22}_{-0.17}$ & $ 9.38$ & $^{+ 8.67}_{- 3.97}$ & $-0.45$ & $^{+0.29}_{-0.32}$ \\
    VCC\,1087 & $ 7.35$ & $^{+ 8.36}_{- 6.34}$ & $-0.67$ & $^{+0.64}_{-0.55}$ & $ 7.49$ & $^{+ 7.73}_{- 6.26}$ & $-0.69$ & $^{+0.52}_{-0.35}$ & $ 7.19$ & $^{+ 8.40}_{- 5.35}$ & $-0.64$ & $^{+0.44}_{-0.47}$ \\
    VCC\,1122 & $ 7.91$ & $^{+ 7.58}_{- 5.95}$ & $-0.87$ & $^{+0.62}_{-0.56}$ & $10.05$ & $^{+10.64}_{- 5.12}$ & $-1.13$ & $^{+0.35}_{-0.48}$ & $ 5.47$ & $^{+ 6.71}_{- 3.41}$ & $-0.46$ & $^{+0.38}_{-0.42}$ \\
    VCC\,1183 & $ 3.79$ & $^{+ 3.18}_{- 1.72}$ & $-0.45$ & $^{+0.49}_{-0.39}$ & $ 4.76$ & $^{+ 3.28}_{- 2.59}$ & $-0.68$ & $^{+0.29}_{-0.31}$ & $ 3.08$ & $^{+ 2.39}_{- 1.12}$ & $-0.21$ & $^{+0.34}_{-0.30}$ \\
    VCC\,1261 & $ 4.00$ & $^{+ 3.33}_{- 1.85}$ & $-0.51$ & $^{+0.49}_{-0.39}$ & $ 5.41$ & $^{+ 3.38}_{- 2.70}$ & $-0.81$ & $^{+0.30}_{-0.32}$ & $ 2.99$ & $^{+ 2.27}_{- 1.03}$ & $-0.17$ & $^{+0.35}_{-0.31}$ \\
    VCC\,1431 & $18.00$ & $^{+61.19}_{-12.20}$ & $-0.82$ & $^{+0.41}_{-0.64}$ & $19.49$ & $^{+57.47}_{- 9.26}$ & $-0.72$ & $^{+0.31}_{-0.71}$ & $18.00$ & $^{+51.18}_{-14.10}$ & $-0.89$ & $^{+0.32}_{-0.37}$ \\
    VCC\,1549 & $12.17$ & $^{+14.11}_{- 7.12}$ & $-0.53$ & $^{+0.50}_{-0.53}$ & $10.68$ & $^{+12.92}_{- 6.75}$ & $-0.35$ & $^{+0.42}_{-0.42}$ & $13.27$ & $^{+46.41}_{- 2.64}$ & $-0.69$ & $^{+0.36}_{-0.42}$ \\
    VCC\,1695 & $ 2.93$ & $^{+ 5.30}_{- 1.09}$ & $-0.72$ & $^{+0.67}_{-0.77}$ & $ 3.49$ & $^{+ 4.67}_{- 1.63}$ & $-0.89$ & $^{+0.49}_{-0.54}$ & $ 2.54$ & $^{+ 3.28}_{- 0.69}$ & $-0.53$ & $^{+0.47}_{-0.54}$ \\
    VCC\,1861 & $ 8.22$ & $^{+24.30}_{- 5.35}$ & $-0.52$ & $^{+0.93}_{-0.89}$ & $ 7.71$ & $^{+22.00}_{- 6.21}$ & $-0.42$ & $^{+0.67}_{-0.91}$ & $ 8.91$ & $^{+25.01}_{- 4.92}$ & $-0.63$ & $^{+0.67}_{-0.80}$ \\
    VCC\,1910 & $ 7.12$ & $^{+ 7.79}_{- 4.30}$ & $-0.25$ & $^{+0.39}_{-0.44}$ & $ 7.41$ & $^{+ 7.93}_{- 4.41}$ & $-0.29$ & $^{+0.33}_{-0.38}$ & $ 6.92$ & $^{+ 6.39}_{- 3.87}$ & $-0.21$ & $^{+0.29}_{-0.27}$ \\
    VCC\,1912 & $ 2.05$ & $^{+ 0.91}_{- 0.54}$ & $-0.92$ & $^{+0.49}_{-0.36}$ & $ 2.50$ & $^{+ 0.79}_{- 0.49}$ & $-1.17$ & $^{+0.25}_{-0.27}$ & $ 1.63$ & $^{+ 0.48}_{- 0.24}$ & $-0.53$ & $^{+0.34}_{-0.32}$ \\
    VCC\,1947 & $ 3.21$ & $^{+ 3.15}_{- 1.47}$ & $ 0.15$ & $^{+0.52}_{-0.41}$ & $ 3.01$ & $^{+ 3.03}_{- 2.07}$ & $ 0.25$ & $^{+0.55}_{-0.48}$ & $ 3.30$ & $^{+ 2.97}_{- 1.25}$ & $ 0.08$ & $^{+0.29}_{-0.27}$ \\
    \hline
    \hline
  \end{tabular}
\end{table*}
\begin{table*}
  \caption{Ages, metallicities and abundance ratios from different indices using TMB03 models} 
  \label{tab_agemet_TMB}
  \begin{tabular}{l*4{r@{}l}}
    \hline
               & \multicolumn{4}{c}{\Hbeta\ -- [MgFe]} & \multicolumn{4}{c}{Mg$b$ -- \Fe} \\
    galaxy     & \multicolumn{2}{c}{age}   & \multicolumn{2}{c}{log(Z/Z$_\odot$)} & \multicolumn{2}{c}{log(Z/Z$_\odot$)} & \multicolumn{2}{c}{[$\alpha$/Fe]} \\
               & \multicolumn{2}{c}{(Gyr)} & \multicolumn{2}{c}{(dex)}            & \multicolumn{2}{c}{(dex)}            & \multicolumn{2}{c}{(dex)}         \\
    \hline
    M\,32     & $ 5.81$ & $^{+ 0.90}_{- 0.79}$ & $-0.10$ & $^{+0.05}_{-0.05}$ & $-0.11$ & $^{+0.02}_{-0.02}$ & $-0.03$ & $^{+0.03}_{-0.03}$ \\
    ID\,0650  & $ 7.34$ & $^{+ 8.85}_{- 3.98}$ & $-0.28$ & $^{+0.29}_{-0.72}$ & $-0.25$ & $^{+0.15}_{-0.18}$ & $-0.19$ & $^{+0.25}_{-0.22}$ \\
    ID\,0734  & $ 6.33$ & $^{+12.02}_{- 4.80}$ & $-0.54$ & $^{+0.59}_{-1.87}$ & $-0.49$ & $^{+0.33}_{-0.46}$ & $-0.23$ & $^{+0.62}_{-0.55}$ \\
    ID\,0872  & $ 3.39$ & $^{+ 3.85}_{- 3.00}$ & $-0.70$ & $^{+0.76}_{-3.16}$ & $-0.61$ & $^{+0.28}_{-0.83}$ & $-0.43$ & $^{+1.21}_{-1.46}$ \\
    ID\,0918  & $ 4.50$ & $^{+ 4.20}_{- 2.11}$ & $-0.12$ & $^{+0.26}_{-0.24}$ & $-0.15$ & $^{+0.08}_{-0.08}$ & $ 0.10$ & $^{+0.13}_{-0.12}$ \\
    ID\,1524  & $ 5.46$ & $^{+ 3.27}_{- 3.00}$ & $-0.96$ & $^{+1.04}_{-2.48}$ & $-0.92$ & $^{+0.48}_{-0.83}$ & $-0.19$ & $^{+1.21}_{-1.35}$ \\
    VCC\,0021 & $ 0.78$ & $^{+ 5.21}_{- 0.52}$ & $-0.71$ & $^{+0.78}_{-1.36}$ & $-0.75$ & $^{+0.41}_{-0.42}$ & $-0.04$ & $^{+0.57}_{-0.57}$ \\
    VCC\,0308 & $ 2.63$ & $^{+ 4.72}_{- 1.71}$ & $-0.25$ & $^{+0.77}_{-0.89}$ & $-0.29$ & $^{+0.23}_{-0.30}$ & $ 0.06$ & $^{+0.50}_{-0.40}$ \\
    VCC\,0397 & $ 1.60$ & $^{+ 3.21}_{- 0.85}$ & $-0.06$ & $^{+0.63}_{-0.70}$ & $-0.11$ & $^{+0.21}_{-0.25}$ & $-0.06$ & $^{+0.28}_{-0.27}$ \\
    VCC\,0523 & $ 3.30$ & $^{+ 3.84}_{- 2.00}$ & $-0.23$ & $^{+0.56}_{-0.71}$ & $-0.22$ & $^{+0.18}_{-0.23}$ & $ 0.19$ & $^{+0.43}_{-0.34}$ \\
    VCC\,0856 & $ 5.90$ & $^{+ 6.76}_{- 3.90}$ & $-0.44$ & $^{+0.56}_{-0.87}$ & $-0.42$ & $^{+0.22}_{-0.29}$ & $ 0.22$ & $^{+0.47}_{-0.36}$ \\
    VCC\,0917 & $ 7.42$ & $^{+ 9.49}_{- 5.00}$ & $-0.48$ & $^{+0.43}_{-0.95}$ & $-0.43$ & $^{+0.20}_{-0.24}$ & $-0.29$ & $^{+0.31}_{-0.30}$ \\
    VCC\,0990 & $11.71$ & $^{+ 6.76}_{- 6.22}$ & $-0.62$ & $^{+0.43}_{-0.69}$ & $-0.58$ & $^{+0.14}_{-0.16}$ & $-0.20$ & $^{+0.23}_{-0.21}$ \\
    VCC\,1087 & $ 7.32$ & $^{+ 9.37}_{- 5.43}$ & $-0.47$ & $^{+0.43}_{-0.95}$ & $-0.45$ & $^{+0.20}_{-0.25}$ & $ 0.12$ & $^{+0.37}_{-0.30}$ \\
    VCC\,1122 & $ 8.01$ & $^{+ 8.32}_{- 5.72}$ & $-0.72$ & $^{+0.55}_{-1.05}$ & $-0.68$ & $^{+0.24}_{-0.28}$ & $-0.33$ & $^{+0.37}_{-0.36}$ \\
    VCC\,1183 & $ 3.48$ & $^{+ 2.54}_{- 1.06}$ & $-0.25$ & $^{+0.36}_{-0.43}$ & $-0.23$ & $^{+0.12}_{-0.14}$ & $-0.05$ & $^{+0.22}_{-0.19}$ \\
    VCC\,1261 & $ 3.74$ & $^{+ 2.35}_{- 2.00}$ & $-0.31$ & $^{+0.38}_{-0.45}$ & $-0.31$ & $^{+0.14}_{-0.16}$ & $-0.15$ & $^{+0.23}_{-0.21}$ \\
    VCC\,1431 & $16.14$ & $^{+ 9.52}_{- 8.07}$ & $-0.63$ & $^{+0.43}_{-0.74}$ & $-0.48$ & $^{+0.12}_{-0.14}$ & $ 0.29$ & $^{+0.23}_{-0.20}$ \\
    VCC\,1549 & $11.55$ & $^{+21.83}_{- 2.41}$ & $-0.30$ & $^{+0.29}_{-0.64}$ & $-0.28$ & $^{+0.13}_{-0.15}$ & $ 0.33$ & $^{+0.28}_{-0.22}$ \\
    VCC\,1695 & $ 2.75$ & $^{+ 4.44}_{- 1.25}$ & $-0.49$ & $^{+0.75}_{-0.86}$ & $-0.51$ & $^{+0.24}_{-0.29}$ & $ 0.03$ & $^{+0.48}_{-0.43}$ \\
    VCC\,1861 & $ 8.47$ & $^{+11.52}_{- 6.31}$ & $-0.31$ & $^{+0.44}_{-1.23}$ & $-0.28$ & $^{+0.23}_{-0.34}$ & $ 0.26$ & $^{+0.58}_{-0.39}$ \\
    VCC\,1910 & $ 7.58$ & $^{+ 9.57}_{- 5.51}$ & $-0.13$ & $^{+0.37}_{-0.26}$ & $-0.13$ & $^{+0.11}_{-0.10}$ & $ 0.05$ & $^{+0.17}_{-0.16}$ \\
    VCC\,1912 & $ 1.35$ & $^{+ 2.00}_{- 0.69}$ & $-0.56$ & $^{+0.42}_{-0.57}$ & $-0.48$ & $^{+0.20}_{-0.20}$ & $-0.03$ & $^{+0.22}_{-0.21}$ \\
    VCC\,1947 & $ 2.99$ & $^{+ 3.68}_{- 0.89}$ & $ 0.19$ & $^{+0.34}_{-0.32}$ & $ 0.20$ & $^{+0.11}_{-0.11}$ & $ 0.12$ & $^{+0.11}_{-0.11}$ \\
    \hline
    \hline
  \end{tabular}
\end{table*}

\begin{figure}
  \includegraphics[clip,width=8cm]{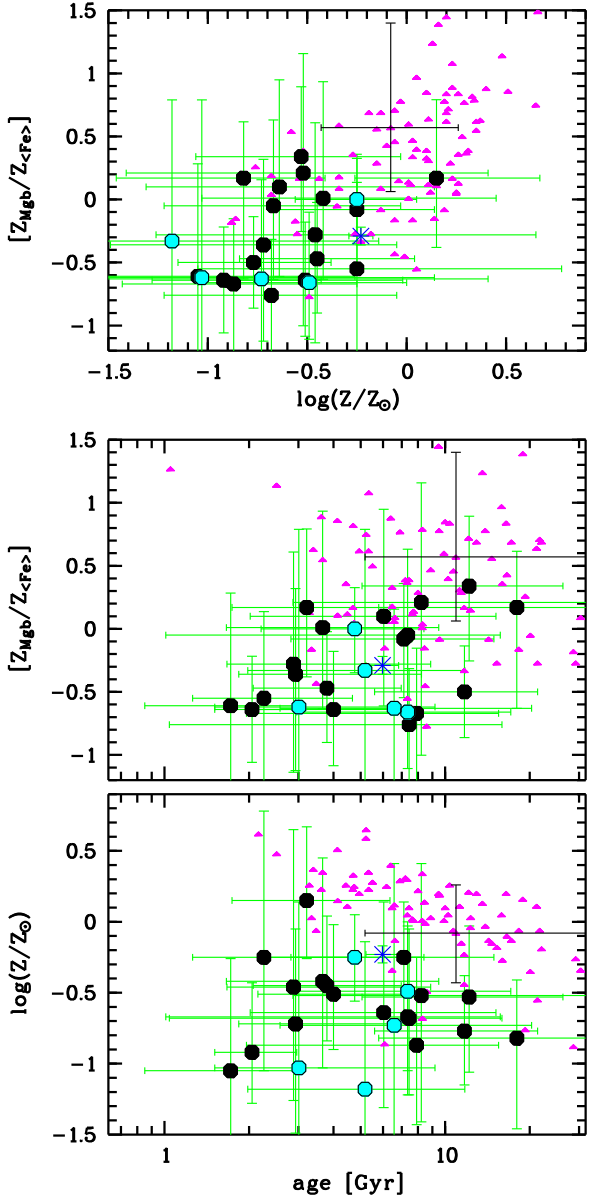}
  \caption{%
    Age versus metallicity, age versus \ZMgZFe\ ratio, and metallicity versus
    \ZMgZFe. The symbols are the same as in
    Figure~\ref{fig_Hbeta_MgFe_our_geha_vzee}. The black error bars are those
    of a typical galaxy in the SB06 sample.}
  \label{fig_agemet}
\end{figure}

\begin{figure}
  \includegraphics[clip,width=8cm]{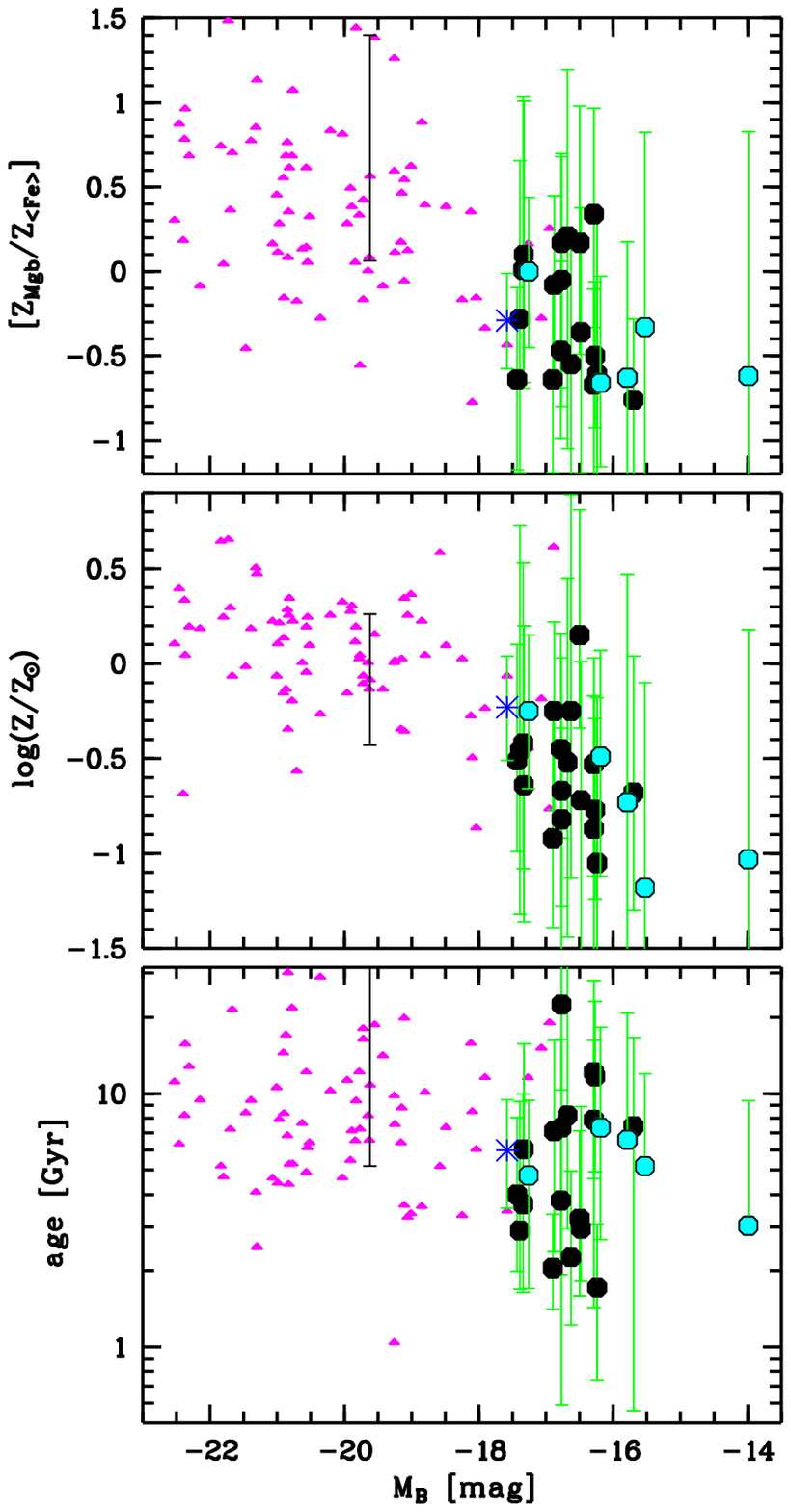}
  \caption{%
    Age, metallicity and \ZMgZFe\ versus $M_B$. Symbols are the same as in
    Figure~\ref{fig_Hbeta_MgFe_our_geha_vzee}.}
  \label{fig_MB_agemet}
\end{figure}
To derive ages and metallicities from measured \Hbeta\ -- [MgFe] indices, we
apply a quadratic interpolation over the 9 nearest SSP model grid points
\citep[see][for a detailed description]{cardieletal03}. To estimate the errors
in the derived quantities we make 1000 Monte Carlo realizations using the
errors on the indices and derive 1$\sigma$ error contours in the age --
metallicity space. These error contours are not necessarily ellipses; their
exact form depends on how the index space maps into the age -- metallicity
space. As a conservative limit, we take the extremes of the error contour as
upper and lower errors on both age and metallicity. In
Table~\ref{tab_agemet_V96} we list the ages and metallicities derived using
\Hbeta\ and different metallicity indicators ([MgFe], Mg$b$ and \Fe) using the
V96 models.  Similarly, using the TMB03 models, we also derive the
[$\alpha$/Fe] abundance ratio and metallicity from the Mg$b$ -- \Fe\ diagrams,
keeping age fixed at the age derived using \Hbeta\ -- [MgFe]. The
metallicities we get from both \Hbeta\ -- [MgFe] and Mg$b$ -- \Fe\ diagrams
are very similar (see Table~\ref{tab_agemet_TMB}).

In Figure~\ref{fig_compare_vaz_TMB} we show the comparison of ages,
metallicities and abundance ratios derived for our sample and the SB06
sample.  Although the V96 models make no predictions for non-solar
[$\alpha$/Fe] ratios, the difference of the metallicities derived from
\Hbeta\ -- Mg$b$ and \Hbeta\ -- \Fe\ diagrams correlates with the
[$\alpha$/Fe] ratio inferred using TMB03 models
\citep[Figure~\ref{fig_compare_vaz_TMB}c, see
also][]{yamadaetal06}. In the following, we will denote this
difference as \ZMgZFe. The ages agree remarkably well
(Figure~\ref{fig_compare_vaz_TMB}a), except for very young systems
($\textrm{age} < 2$\,Gyr), where the TMB03 models give slightly
younger ages. For the metallicity, there is a systematic 0.2 dex
offset towards higher metallicities in TMB03
(Figure~\ref{fig_compare_vaz_TMB}b); this offset seems to become
slightly larger when going to lower metallicities. Finally, \ZMgZFe\
appears to be a good proxy for the [$\alpha$/Fe] abundance ratio
(Figure~\ref{fig_compare_vaz_TMB}c, see also \citet{delarosa07}).  In
the following we will use the ages, metallicities and \ZMgZFe\ derived
using V96 models, but the choice of models does not alter the
conclusions whatsoever.

The dEs all fall inside the model grids. However, many of the Es have index
measurements that lie outside the region defined by the models. In those
cases, the derived ages/metallicities are extrapolations outside the model
grids and should be treated with caution. For a more detailed examination of
the ages and metallicities of the Es, we refer the reader to
\citet{sanblasetal06b}.

In order to estimate the robustness of the ages and metallicities derived, we
compare the behaviour of different age and metallicity indicators
(Figure~\ref{fig_compare_agemet}). The ages of the dEs, derived from different
Balmer indices, are consistent, but with a slight hint that \HgF\ 
overestimates the age of the dEs. The ages of the Es however, are
underestimated by both \HgF\ and \HdF, a known effect of their super-solar
[$\alpha$/Fe] abundance ratios \citep{thomasetal04}. This effect is even more
noticeable in the derived metallicities (Mg$b$ gives higher metallicities,
\Fe\ gives lower metallicities). The metallicity of the dEs however is
consistent, irrespective of the metallicity index used. Although the error
bars are large, the effect of the sub-solar [$\alpha$/Fe] ratio is present for
the field dEs; Mg$b$ gives slightly lower metallicities, \Fe\ gives slightly
higher metallicities.

In Figure~\ref{fig_agemet} we show the derived age versus metallicity and
\ZMgZFe, and metallicity versus \ZMgZFe. Although the index -- index diagrams
are not completely orthogonal in age, metallicity and abundance ratio, the
idea that dEs are on the whole younger and less metal-rich, and that they have
lower abundance ratios than Es, is confirmed. The age -- metallicity plot
shows the extension of the \citet{trageretal00} age -- metallicity -- velocity
dispersion projection. For galaxies with the same velocity dispersion, age and
metallicity anti-correlate, but the lower the velocity dispersion, the lower
the mean ages and metallicities. This point will be explored in greater detail
in \citet{toloba07} when we have the velocity dispersions for our dEs.

Using the whole dE sample and the V96 model predictions, we find a
mean age of $6.1 \pm 3.8$\,Gyr and a mean metallicity of
log(Z/Z$_\odot$)$ = -0.60 \pm 0.31$. For only the Virgo sample we find
an age of $6.2 \pm 4.3$\,Gyr and a metallicity of log(Z/Z$_\odot$)$ =
-0.58 \pm 0.28$. For only the Virgo sample, and using the TMB03
models, we find a mean age of $5.9 \pm 4.0$\,Gyr and mean metallicity
of log(Z/Z$_\odot$)$ = -0.38 \pm 0.24$, in very good agreement with
the results of \citet{gehaetal03}. They find, for a sample of 17 Virgo
Cluster dEs and also using TMB03 models, a mean age of $5 \pm 2$\,Gyr
and mean metallicity of log(Z/Z$_\odot$)$ = -0.3 \pm 0.1$,
respectively.

Although the mean age found for our dE sample does not change much if
we include or exclude the field dEs, it is interesting to see that
none of the field dEs has an age larger than 8\,Gyr. On the other
hand, there are also no field dEs younger than 3\,Gyr, while some
Virgo dEs are as young as 2\,Gyr. As a statistical comparison, we use
the one-dimensional Kolmogorov-Smirnov (K-S) test. This test gives the
probability ($P_{\rm KS}$) that the difference between two
distributions would be as large as observed if they had been drawn
from the same population, and works well even for small samples. We
have to take into account that age, metallicity and abundance ratios
are determined from the \Hbeta, Mg$b$ and \Fe\ index data sets. The
derived quantities are obviously correlated because the model grids
are not orthogonal in the index space. Therefore we perform the K-S
test on the measured indices rather than on the derived quantities.
We find for the \Hbeta, Mg$b$ and \Fe\ distributions a $P_{\rm KS}$
value of 0.68, 0.96 and 0.52, respectively.  Therefore we cannot rule
out that the indices of field and Virgo dEs have the same
distribution.

In Figure~\ref{fig_MB_agemet} we show $M_B$ versus age, metallicity and
\ZMgZFe\ for the galaxies in our sample and those in the sample from
SB06 with $M_B$ available in HYPERLEDA (Table~1 in SB06). There is a clear
correlation between $M_B$ and age, metallicity and abundance ratio. In all
cases, dEs form the low-mass tail of the correlations for Es.  In age, the dEs
are generally younger than the Es. In metallicity, the dEs extend the
luminosity-metallicity relation towards lower luminosities. Finally, the dEs
have lower abundance ratios than the massive Es.

The high-metallicity dE is VCC\,1947, which is known to rotate. This could be
evidence that VCC\,1947 stems from a harassed, more massive spiral.  However,
this dE has also been observed by \citet{gehaetal03} who find a lower
metallicity, so we should be careful with this galaxy.  Moreover, other
rotationally supported dEs in our sample, such as VCC\,397 and VCC\,1122, or
VCC\,856 which has a spiral structure, do not show such a high metallicity.

\subsection{Stellar light distributions}
\label{sec_lightdistribution}

\begin{table}
  \caption{Structural parameters $C$, $A$, $S$}
  \label{tab_cas}
  \begin{tabular}{l*3{r@{ $\pm$ }l}}
    \hline
    galaxy     & \multicolumn{2}{c}{$C$} & \multicolumn{2}{c}{$A$} & \multicolumn{2}{c}{$S$} \\
    \hline    
    M\,32     & \multicolumn{2}{c}{---} & \multicolumn{2}{c}{---} & \multicolumn{2}{c}{---} \\
    ID\,0650  &3.881 & 0.028 & 0.023 & 0.012 & $-$0.26 & 0.36 \\
    ID\,0734  &2.399 & 0.012 & 0.038 & 0.015 &    0.33 & 0.33 \\
    ID\,0872  &3.768 & 0.055 & 0.053 & 0.019 & $-$0.23 & 0.42 \\
    ID\,0918  &4.145 & 0.087 & 0.048 & 0.002 &    0.07 & 0.06 \\
    ID\,1524  &2.293 & 0.013 & 0.133 & 0.010 &    0.21 & 0.39 \\
    VCC\,0021 &3.128 & 0.045 & 0.147 & 0.008 &    0.11 & 0.14 \\
    VCC\,0308 &2.930 & 0.033 & 0.058 & 0.012 &    0.07 & 0.11 \\
    VCC\,0397 &2.967 & 0.051 & 0.112 & 0.021 &    0.06 & 0.09 \\
    VCC\,0523 &2.761 & 0.034 & 0.059 & 0.008 &    0.02 & 0.05 \\
    VCC\,0856 &2.561 & 0.035 & 0.119 & 0.008 &    0.09 & 0.11 \\
    VCC\,0917 &3.266 & 0.060 & 0.101 & 0.008 &    0.07 & 0.09 \\
    VCC\,0990 &3.202 & 0.054 & 0.022 & 0.007 &    0.04 & 0.05 \\
    VCC\,1087 &2.877 & 0.033 & 0.041 & 0.010 & $-$0.06 & 0.10 \\
    VCC\,1122 &3.088 & 0.043 & 0.062 & 0.007 &    0.09 & 0.11 \\
    VCC\,1183 &3.204 & 0.040 & 0.055 & 0.016 &    0.11 & 0.11 \\
    VCC\,1261 &2.991 & 0.036 & 0.084 & 0.007 & $-$0.02 & 0.04 \\
    VCC\,1431 &2.959 & 0.048 & 0.034 & 0.006 &    0.02 & 0.04 \\
    VCC\,1549 &3.170 & 0.039 & 0.104 & 0.005 &    0.12 & 0.20 \\
    VCC\,1695 &3.476 & 0.036 & 0.048 & 0.009 &    0.12 & 0.11 \\
    VCC\,1861 &3.107 & 0.029 & 0.001 & 0.015 &    0.18 & 0.22 \\
    VCC\,1910 &3.079 & 0.039 & 0.091 & 0.003 &    0.11 & 0.16 \\
    VCC\,1912 &3.485 & 0.044 & 0.128 & 0.007 &    0.06 & 0.08 \\
    VCC\,1947 &2.991 & 0.050 & 0.079 & 0.005 &    0.05 & 0.08 \\
    \hline
    \hline
  \end{tabular}
\end{table}

\begin{figure}
  \includegraphics[clip,width=8cm]{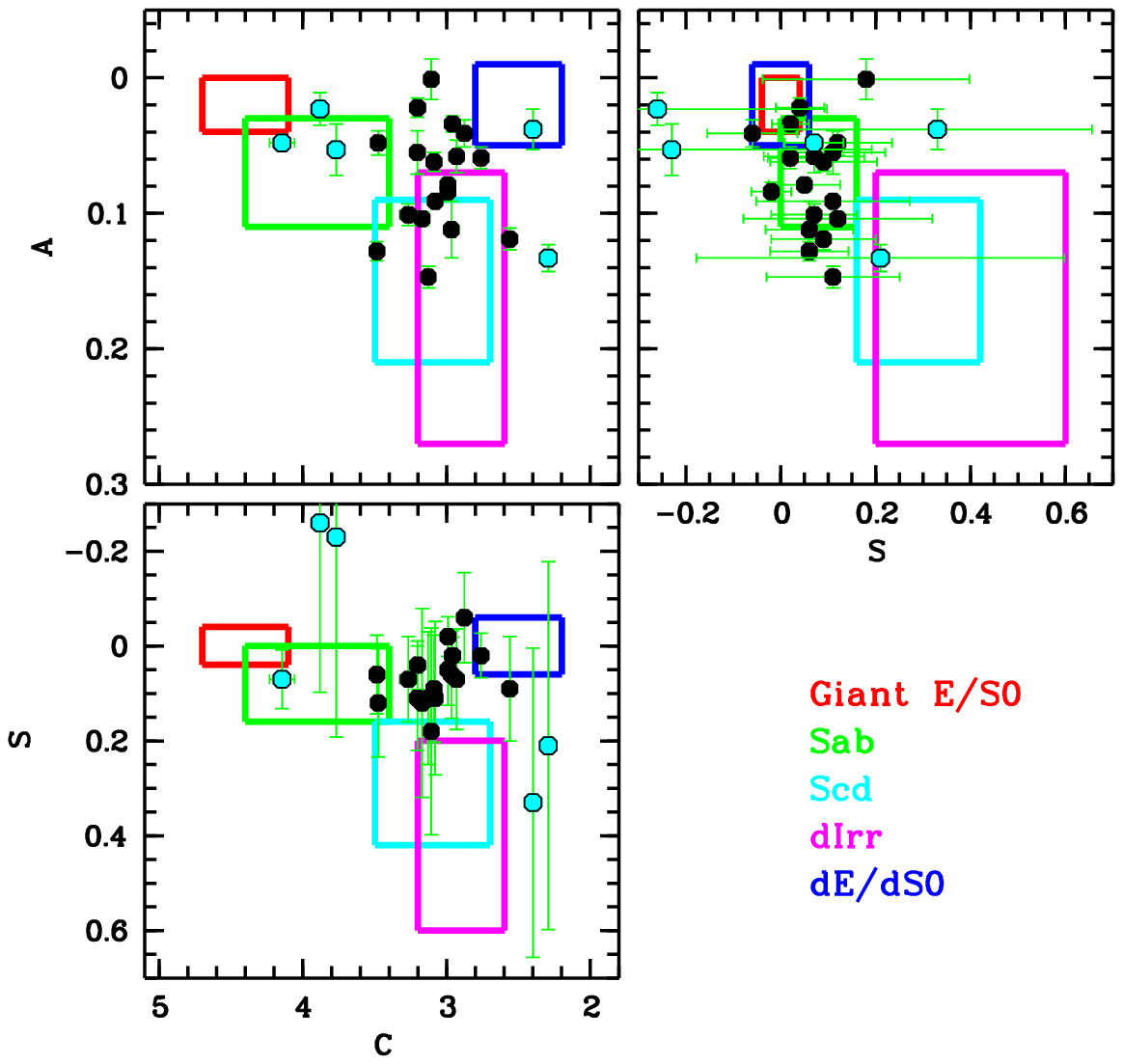}
  \caption{%
    Structural parameters ($C$, $A$ and $S$, see text) for the dEs in our
    sample. Cyan-filled symbols are field dEs, solid symbols are Virgo dEs.
    The locus of different galaxy types are indicated. Note the selection
    effects on the field dEs. We either select quite diffuse or quite
    concentrated galaxies.}
  \label{fig_cas}
\end{figure}

Using the SDSS $g$-band images, we analyse the structural parameters of the
galaxies in our sample. The concentration ($C$), large-scale asymmetry ($A$)
and clumpiness ($S$), are three model-independent parameters that can be used
to quantify a galaxy's structural appearance \citep{conselice03}.

The CAS parameters have a well-defined range of values and are
computed using simple techniques.  The concentration index is the
logarithm of the ratio of the radius containing 80\%\ of the light in
a galaxy to the radius which contains 20\%\ of the light
\citep{conselice00}. The range in $C$ values is found from 2 to 5,
with higher $C$ values for more concentrated galaxies, such as massive
early types. The asymmetry is measured by rotating a galaxy's image by
$180\deg$\ and subtracting this rotated image from the original
galaxy's image.  The residuals of this subtraction are compared with
the original galaxy's flux to obtain a ratio of asymmetric light.  The
radii and centering involved in this computation are well-defined and
explained in \citep{conselice00}. The asymmetry ranges from 0 to $\sim
1$ with merging galaxies typically found at $A > 0.35$. The clumpiness
is defined in a similar way to the asymmetry, except that the amount
of light in high frequency 'clumps' is compared to the galaxy's total
light \citep{conselice03}.  The $S$ values range from 0 to $> 2$, with
most star forming galaxies have $S > 0.3$.

In Figure~\ref{fig_cas}, we show the values for $C$, $A$ and $S$ measured on
the whole sample (see also Table~\ref{tab_cas}). The range adopted for the
$C$, $A$ and $S$ plots span the range measured for different galaxy types,
taken from Table~6 of \citet{conselice03}. The dEs in the sample used in that
work have lower luminosities ($M_B = -14.2 \pm 0.9$) than the dEs in our
sample ($M_B = -16.9 \pm 0.9$). The locus of our dEs coincides with what one
expects for these (by selection), smooth, symmetric, diffuse galaxies,
in-between the early-type Es/S0s and the fainter dEs studied in
\citet{conselice03}. A Spearman rank-order test on the whole sample showed
that no significant correlations exist between the $C$, $A$ or $S$ parameters.
However, the errors for the field dEs, especially for clumpiness ($S$) are
quite large.  Using only the Virgo dEs, a correlation between concentration
($C$) and clumpiness ($S$) exists at the 97.5\% confidence level.

We find that the field dEs we have selected are either more concentrated or
less concentrated than the Virgo dEs. This might be a result of the difficulty
in finding dEs in the field, favouring quite compact or very diffuse systems
to be selected. The K-S test yields that the probability that $C$ follows the
same distribution for field and Virgo dEs is 1\% ($P_{\rm KS} = 0.01$), thus
the field and Virgo dEs have a significantly different distribution in
concentration. For $A$ and $S$, $P_{\rm KS}$ gives 0.52 and 0.13,
respectively, so we cannot definitely say they are drawn from a different
distribution.

\begin{figure*}
  \includegraphics[clip,width=16cm]{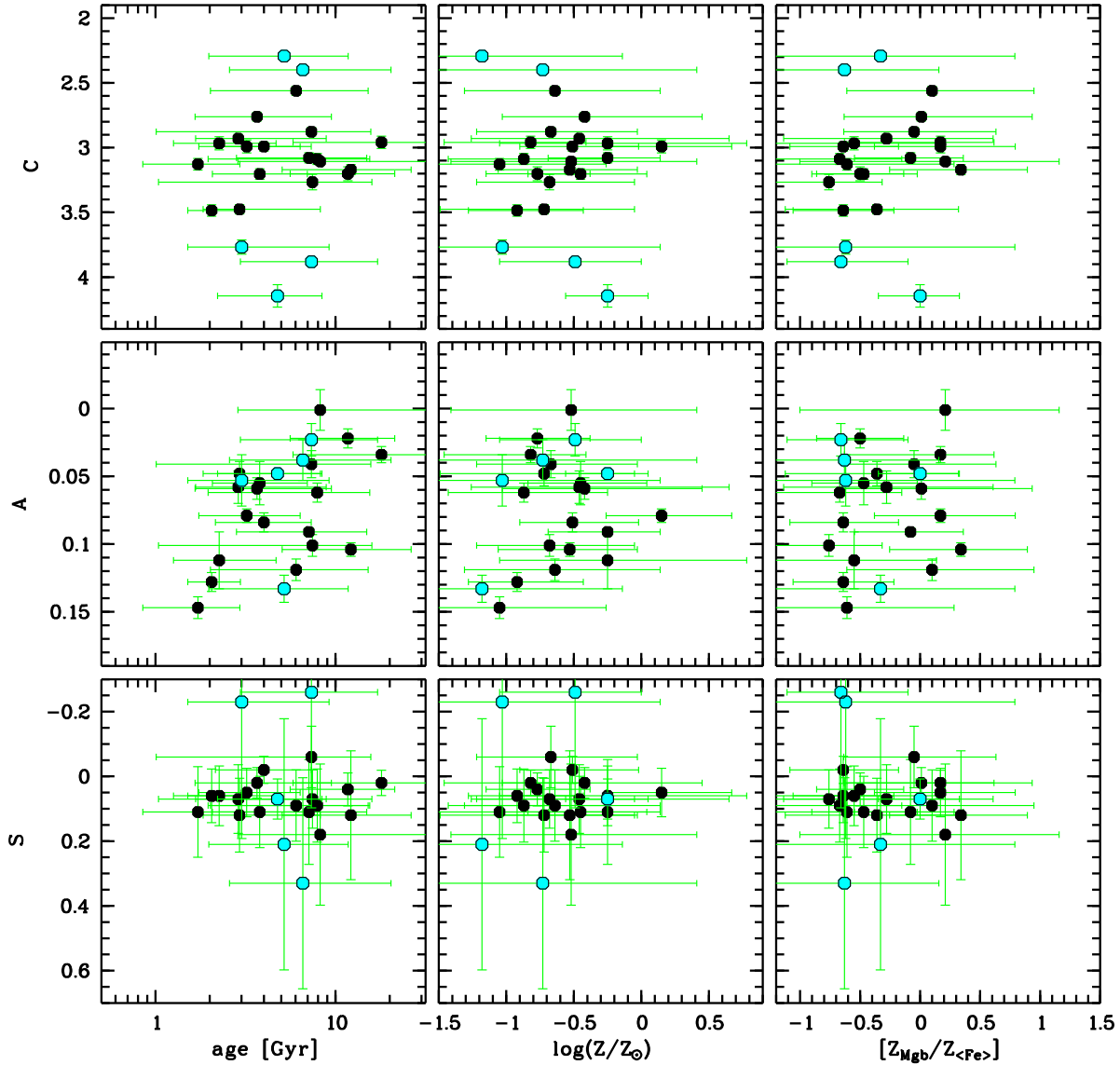}
  \caption{%
    Structural parameters ($C$, $A$ and $S$) of the dEs versus age,
    metallicity and \ZMgZFe. The only significant correlation is
    between log(age) and asymmetry $A$, indicating a connection between the
    dynamical state of the galaxy and its age.}
  \label{fig_cas_agemet}
\end{figure*}

In Figure~\ref{fig_cas_agemet}, we show $C$, $A$ and $S$ as a function of age,
metallicity and \ZMgZFe. Here, a Spearman rank-order test reveals that there
exists a significant (anti)correlation between age and large-scale asymmetry
($A$), at the 97.5\% confidence level using only the Virgo data, and better
than 99\% using the whole sample. Since $A$ measures the large-scale or bulk
asymmetry, it appears that at the same time star formation was switched off,
the galaxy also received a dynamical disturbance leaving an imprint in the
large-scale structure of that galaxy, consistent with a scenario that the
young dEs have only recently fallen into the cluster.

Using only the Virgo sample, it seems there are anti-correlations between
concentration $C$ and metallicity ($>$95\%), and between concentration $C$ and
\ZMgZFe\ ($>$97.5\%). We also find that for the Virgo galaxies, the
concentration parameter $C$ also anti-correlates with $M_B$.  Thus, the found
anti-correlations with $C$ can be traced back to the mass -- metallicity and
mass -- \ZMgZFe\ relations. These correlations are in accordance
with the findings of \citet{vazdekisetal04}, who demonstrated that [Mg/Fe]
correlates stronger with S\'{e}rsic-$n$ than metallicity. Like these authors,
we also find that the correlation with metallicity as estimated by Mg$b$ is
stronger, while it disappears if metallicity estimated by \Fe\ is used. We
will investigate this matter in more detail in subsequent papers, using
photometric and kinematical data.

Finally, because the dEs are selected to be non-starforming systems, and the
clumpiness parameter $S$ correlates very well with the H$\alpha$ emission, we
expect all our dEs to be smooth and to not show a large variation in $S$.

\subsection{The Virgo sample}
\label{sec_virgodistance}

\begin{figure*}
  \includegraphics[clip,width=16cm]{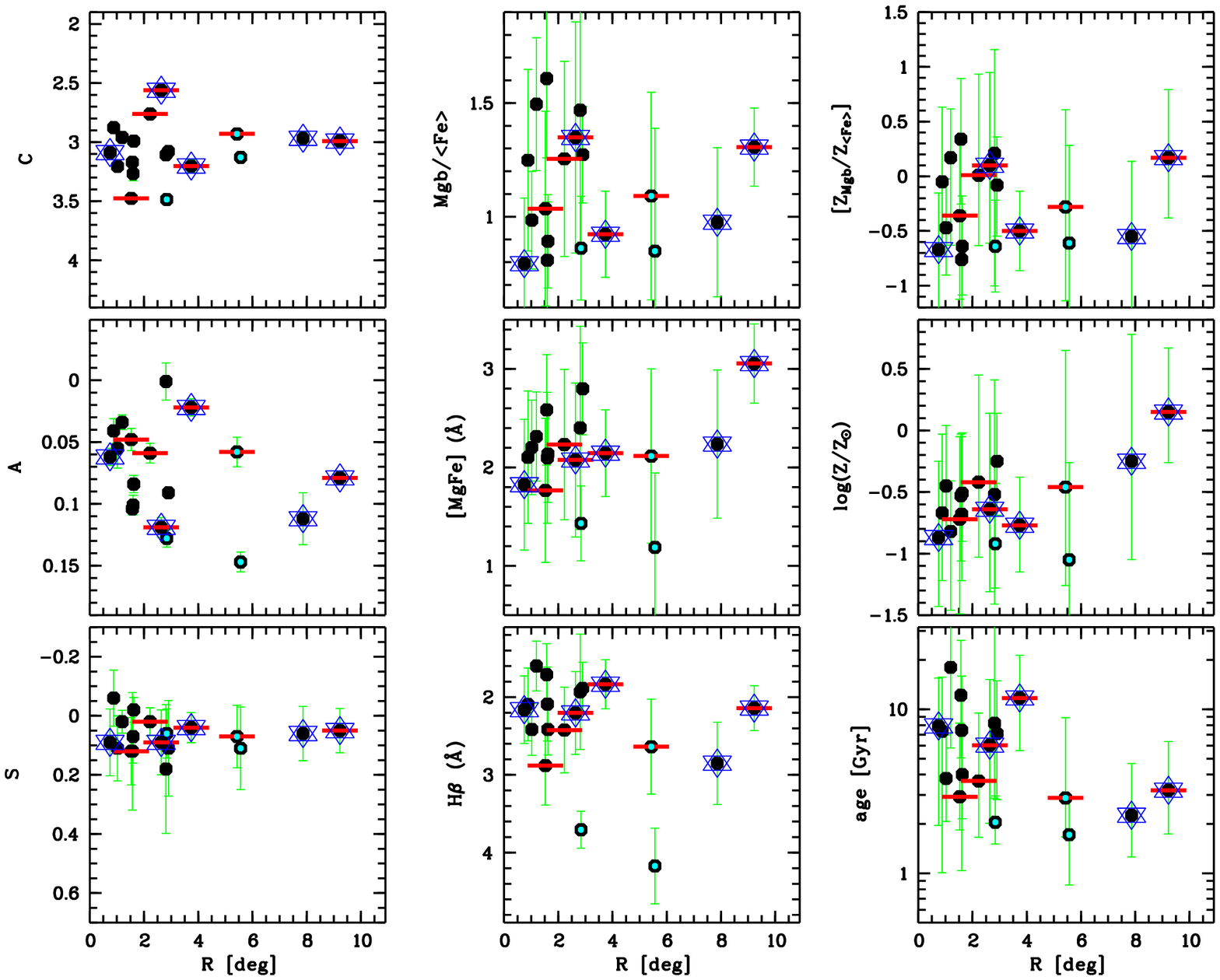}
  \caption{%
    Structural parameters ($C$, $A$ and $S$), measured \Hbeta, [MgFe] and
    Mg$b$/\Fe\ indices, and derived ages, metallicity and \ZMgZFe\ versus the
    projected Virgocentric distance ($R$, in degrees). The symbols with a cyan
    centre are dEs with a blue nucleus \citep{liskeretal06bluenuc}.  Those
    with a red horizontal bar are dEs in which disk or spiral structures have
    been found \citep[certain and probable disks from][]{liskeretal06disk}.
    Finally the stars highlight those dEs with known rotation
    \citep{gehaetal03, vanzeeetal04, toloba07}.}
  \label{fig_virgo_radius}
\end{figure*}

In Figure~\ref{fig_virgo_radius}, we plot the measured $C$, $A$ and
$S$ parameters, the measured \Hbeta, [MgFe] and Mg$b$/\Fe\ indices and
the derived ages, metallicities and \ZMgZFe\ versus the projected
Virgocentric distance (we take M87 as the cluster centre). The
Spearman rank-order test suggests a trend between $R$ and $A$ (better
than 90\%). The $R - A$ trend indicates the effect of the cluster on
the dynamical state of dEs. Also age, and to a lesser extent, \ZMgZFe\
are correlated with the distance to the cluster centre. The Spearman
rank-order significance for the $R$ -- log(age) correlation is better
than 97.5\%. The young dEs lie towards the outskirts of the cluster,
and old dEs towards the centre. Although the Spearman rank-order test
gives low significance to an $R$ -- \ZMgZFe\ correlation, it seems
that the dEs with higher abundance ratio are located in the central 3
degrees (this is a consequence of the age -- \ZMgZFe\ correlation).

We also indicate in Figure~\ref{fig_virgo_radius} those dEs that have
blue nuclei \citep{liskeretal06bluenuc}, disk or spiral structures
\citep[certain and probable disks from][]{liskeretal06disk} and
rotation \citep{gehaetal03, vanzeeetal04, toloba07}. The fraction of
such dEs with residual structure decreases in the centre of the
cluster (less than 2 degrees away from M87), again indicating the
impact of the environment on those low-mass systems.

\section{Discussion}
\label{sec_discussion}

\subsection{Field and Virgo dEs}

This is the first dE study to include a detailed investigation of field
systems.  It is notoriously difficult to find field non-starforming dwarfs,
which in itself already reveals the importance of the environment to stop star
formation, at least in low-mass objects. Except for ID~872, which is a quite
faint dE, the field sample has similar absolute blue magnitudes as the Virgo
dEs, so we are probing a similar mass range. Although the sample selection was
mostly based on colour cuts, the analysis of the light distributions
(Section~\ref{sec_lightdistribution}) reveals that the field galaxies selected
have a different concentration distribution than the Virgo dEs, being either
more extended or more compact than the Virgo cluster dEs of which all but two
are nucleated.

Although our field sample is small (only 5 galaxies) there appear to
be no systematic differences between the Lick indices or the ages,
metallicities and abundance ratios for the Virgo and the field
samples. The K-S test cannot rule out that the samples are drawn from
the same population. Having said that, none of the field dEs has a
luminosity-weighted age larger than 8\,Gyr, whereas the some of Virgo
dEs have ages up to 18\,Gyr. One could argue that also none of the
field dEs is older then 3\,Gyr, while some Virgo dEs are as young as
2\,Gyr. Although the field sample suffers from poor statistics on both
sides, the sample selection based on finding red objects should be
biased towards old systems. Moreover, we found that three of the field
dEs in the original sample are apparently still actively starforming.

It seems that by selecting red, smooth dwarf galaxies (dEs), we are
automatically selecting galaxies with the same properties. It would be
interesting to study the properties of dwarf galaxies \textit{irrespective of
  classification}, as a function of environment. Given that dEs span such a
large range in ages, and dIrrs are still starforming, and given that optical
magnitudes are sensitive to age, it seems that selection should better be done
on the near-infrared magnitudes in order to study connections between
different populations of dwarf galaxies. Indeed, the $H$-band surface
brightness profiles of peculiar dEs and dIrrs appear to be indistinguishable
\citep{gavazzietal01}, and optical structural parameters for dEs and dIrrs are
quite similar as well \citep{vanzeeetal04}.  In subsequent papers from the
MAGPOP-ITP we will address this issue; see also \citet{boselli07}.

\subsection{The evolution of dEs}

\subsubsection{Internal mechanisms}

From this and previous studies, it is becoming clear that not all dEs
are old, primordial objects, and that some of them formed stars until
auite recently ($\sim 2$\,Gyr ago).

It appears that, like for late-type galaxies \citep{bosellietal01},
the star formation history of quiescent galaxies is a function of
mass. Massive early-type galaxies form early and on short time-scales
(with high star formation efficiency), whereas less massive early-type
galaxies have more extended star formation histories (lower star
formation efficiency), leading to (sub-)solar [$\alpha$/Fe] abundance
ratios and young luminosity-weighted ages.  This dependence of star
formation duration and star formation efficiency on mass is reproduced
in detailed N-body/SPH simulations of isolated galaxies
\citep{carraroetal01}.

A dwarf galaxy will only be classified as a dE once it stops star formation
either through exhausting its gas, or through blowing it out in a galactic
wind. Galactic winds, if present, are probably not efficient in blowing away
all the gas, even from low-mass objects \citep{maclowferrara99}. They may
however, preferentially blow away the ejecta from supernova type~II, which are
linked to the sites of star formation and therefore occur quite concentrated
in space and time, while the ejecta from supernova type~Ia are mixed into the
interstellar medium more easily as they occur only sporadically
\citep[e.g.][]{vader86}. This could account for the sub-solar abundance ratios
observed in some of the dEs. Abundance ratios for starforming dwarfs are
unfortunately not yet available. In starforming systems optical emission lines
are present, making the analysis of the underlying stellar populations
difficult. However it seems that the same trends with mass are also found for
late-type spirals, with lower [$\alpha$/Fe] abundance ratios for later Hubble
types \citep{gandaetal07}.

\subsubsection{External mechanisms}

The observed correlation between age and Virgocentric distance indicates that
environment also plays an important role in the evolution of dwarf galaxies.
Similar trends of age and also [$\alpha$/Fe] were observed by
\citet{smithetal06}, who point out that: ``Further progress in this area will
be driven by improved spectroscopic observations of faint cluster members,
which appear to exhibit stronger signatures of later accretion.''

Simulations predict that gas removal by ram pressure stripping in an
intragroup or intracluster medium is very efficient, and proceeds in a few 100
Myr, even in low-density group environments \citep{moriburkert00,
  marcolinietal03}. The morphological transformation of a disk or irregular
galaxy into a more spheroidal, relaxed dE through interactions with other
galaxies and the cluster potential may take longer, up to a few Gyr
\citep{mooreetal98}. Both these environmental effects would leave their
imprint on the galaxies, either in their stellar populations because the star
formation is stopped earlier than in an isolated environment, or in their
stellar light distributions if the interaction has a dynamical effect.

Using our results, we can disentangle the effect of the two mechanisms at
work. The correlation of age with Virgocentric distance, and the fact that
very few genuine intermediate-type dE/dIrr galaxies exist, points to rapid
loss of gas and subsequent truncation of the star formation once a dwarf
galaxy enters the cluster. The correlation of age and bulk asymmetry shows the
morphological transformation at work in galaxies that already stopped star
formation some time ago. The dependency of asymmetry on Virgocentric distance
might be less strong than the correlation with age because it can take several
cluster crossing times\footnote{The crossing time of the Virgo cluster is
  about one-tenth of the Hubble time or slightly more than 1\,Gyr
  \citep{threnthamtully02}.} to complete the morphological transformation.

This is corroborated by the fact that Virgo dEs with blue nuclei, residual
disks, rotation, etc. (see Figure~\ref{fig_virgo_radius}), tend to lie at
larger radii. The spatial distribution of such 'special' dEs appears to be
consistent with that of star forming dwarfs \citep{liskeretal07}, pointing
towards an ongoing infall and transformation of star forming dwarfs into
quiescent dEs in the Virgo cluster. Given the continuous change of age with
Virgocentric distance, and the fact that all but two of the Virgo dEs are
nucleated and relatively bright, it seems that 'normal' and 'special' dEs are
not two subclasses but rather form a continuum of increasingly older and more
relaxed galaxies as they have spend more time in the cluster.

So if star formation is stopped by ram pressure stripping, why did field dE
stop forming stars? We may see the field dEs in a quiescent stage of their
life. Observations of the Local Group dEs reveal that their star formation
history is episodic with gaps of up to a few Gyr \citep[e.g.][]{grebeletal03}.
Studies of HI find that gas-rich dEs, which are mainly located in the
outskirts of clusters and in groups, have gas mass fractions comparable to
those of star forming galaxies \citep{conseliceetal03HI, buyleetal05,
  bouchardetal05}. Some of these dEs even show evidence of ongoing
star-formation at a very low rate \citep{derijckeetal03ism,michielsenetal04}.
Radio observations of the neutral gas content of the field dEs would be a
valuable test of this idea.

\section{Conclusions}
\label{sec_conclusions}

By analysing the stellar populations of a sample of 18 dEs in the Virgo
cluster and 5 field dEs + M\,32 we discover a relationship between the ages of
the stellar populations in dwarfs, their environment and structure. Our
results can be summarised as follows:

\begin{itemize}
\item Unlike massive Es, the [$\alpha$/Fe] abundance ratios of dEs
  scatter around solar, some have even sub-solar abundance
  ratios. This points to an extended or burst-like star formation
  history in dEs, similar to what is found in the Local Group
  dEs. Interestingly, dEs also exhibit different C and N abundance
  ratios than massive Es and globular clusters.
  
\item On average, dEs are younger and less metal-rich than more massive Es, in
  accordance with the 'downsizing' scenario.

\item Although our sample of field dEs is small, there is no statistical
  evidence that the distribution in age, metallicity or abundance ratio is
  different from the Virgo sample. This implies that the chemical evolution of
  dEs is an internally governed process of slow self-enrichment. However,
  preliminary truncation of the star formation by a hostile environment can
  stop this process.
  
\item There are no very old field dEs, and we find that age is
  correlated with projected distance to the Virgo Cluster centre,
  indicating that the cluster environment plays an important role in
  the evolution of dEs through the truncation of star formation,
  probably via ram pressure stripping.

  
\item From the analysis of the structural parameters of the dEs, we show that
  the (mean, luminosity-weighted) age and the bulk asymmetry are correlated.
  The younger dEs show higher internal bulk large-scale distortions. If dEs
  stem from a progenitor population of star forming irregular or disk galaxies
  that quickly stopped star formation after entering the cluster environment
  through ram-pressure stripping and subsequent slow transformation to more
  spheroidal objects through harassment, we indeed expect those dEs that fell
  in early to be more relaxed and symmetric than those that were accreted more
  recently.

\end{itemize}

In subsequent papers from the MAGPOP-ITP, we will investigate the kinematics
of this sample to compare their stellar versus dynamical mass-to-light ratios
and their place in the fundamental plane \citep{toloba07}.

\section*{Acknowledgements}

We would like to thank Scott Trager, Ignacio Trujillo and an anonymous
referee for valuable comments and discussion. DM thanks the MAGPOP EU
Marie Curie Training and Research Network for financial support.  The
Network also provided financial support for collaborating research
visits during which part of this work was done. Based on observations
made with the NOT, operated on the island of La Palma jointly by
Denmark, Finland, Iceland, Norway, and Sweden, in the Spanish
Observatorio del Roque de los Muchachos of the Instituto de
Astrof\'{i}sica de Canarias.  The data presented here have been taken
using ALFOSC, which is owned by the Instituto de Astrof\'{i}sica de
Andaluc\'{i}a (IAA) and operated at the NOT under agreement between
IAA and the NBIfAFG of the Astronomical Observatory of Copenhagen.
This paper made use of the following public databases: SDSS, NED,
HyperLEDA, GOLDMine.

\appendix

\section{Transformation to the Lick/IDS system}
\label{app_transformation}

\begin{figure*}
  \includegraphics[clip,width=\textwidth]{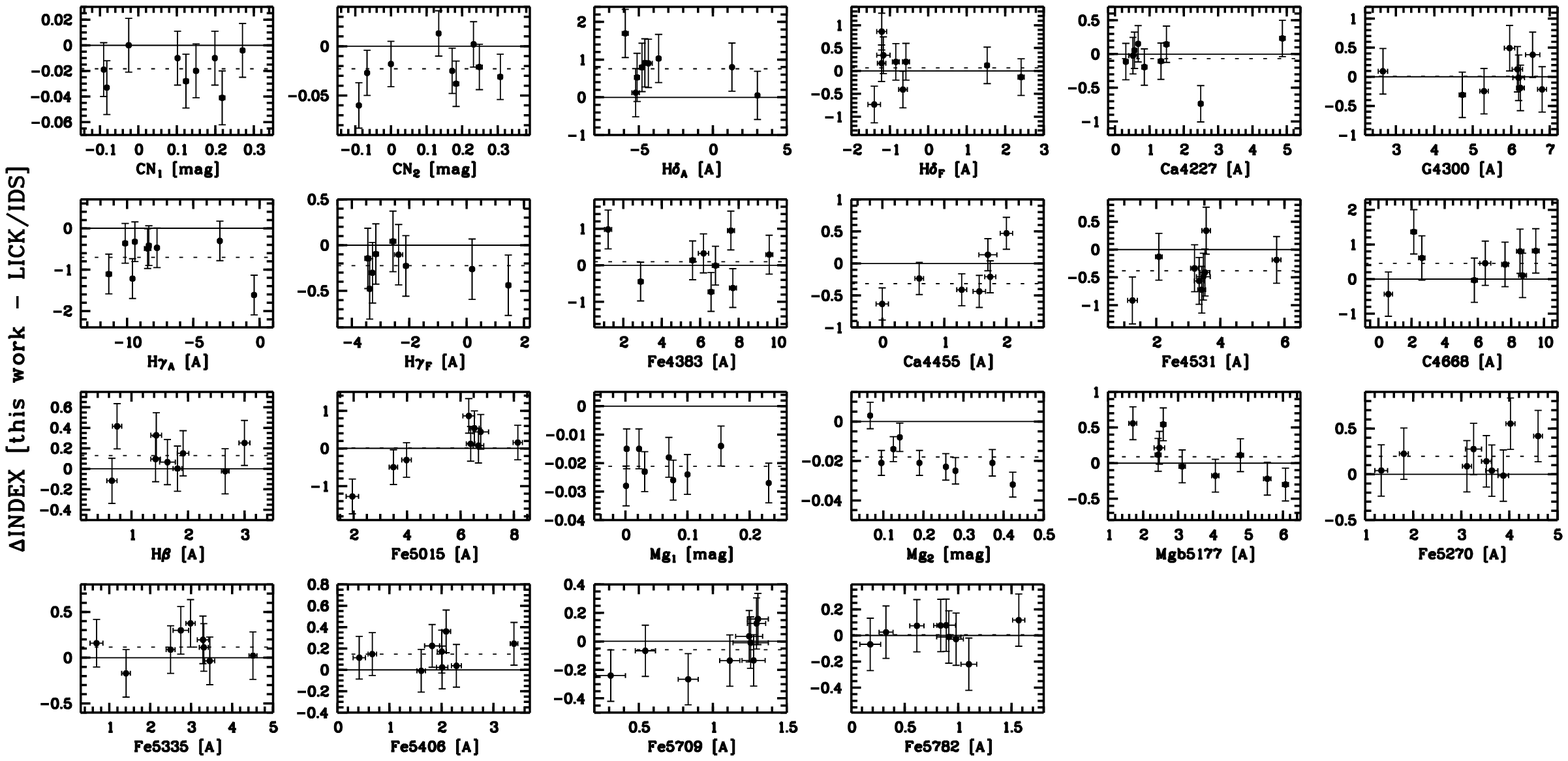}
  \caption{Offsets between the indices measured in the Lick/IDS
    stars and our own measurements (in function of our
    measurements). The dotted line indicates the mean offset.}
  \label{fig_lickoffsets}
\end{figure*}

\begin{table}
  \begin{center}
  \caption{Standard stars}
  \label{tab_standardstars}
  \begin{tabular}{llll}
    \hline
    star & type & MILES & Lick/IDS \\
    \hline
    HD060522 & M0 III & yes & yes \\
    HD065900 & A1 V   & yes & no  \\
    HD072184 & K2 III & yes & yes \\
    HD072324 & G9 III & yes & yes \\
    HD074377 & K3 V   & yes & yes \\
    HD074442 & K0 III & yes & yes \\
    HD075732 & G8 V   & yes & yes \\
    HD085235 & A3 IV  & yes & no  \\
    HD137471 & M1 III & yes & yes \\
    HD140160 & Aop V  & yes & no  \\
    HD143761 & G2 V   & yes & yes \\
    HD144872 & K3 V   & yes & yes \\
    HD148513 & K4 III & yes & yes \\
    HD165760 & G8 III & no  & yes \\
    HD165908 & F7 V   & yes & yes \\
    \hline
    \hline
  \end{tabular}
  \end{center}
\end{table}
During the course of the observations, we observed 15 standard stars
(see Table~\ref{tab_standardstars}). Fourteen of stars are in the
(relative) flux-calibrated MILES sample \citep{sanblas06miles} and
were used to obtain a more robust solution for the flux
calibration. We have also 12 stars in our sample that appear in the
original Lick/IDS stellar library \citep{wortheyetal94}. Those have
spectral type later than F and were used to calculate the offsets
between our flux-calibrated spectra and the Lick/IDS response
function. For each index we broadened our spectra to a specific
resolution, as indicated in Table~\ref{tab_offsets} \citep[taken
from][]{gorgasetal07}. Three of the stars had different offsets than
the other 9. It results that these three stars (HD\,074377,
HD\,137471, and HD\,148513) were observed during a very early run of
the Lick/IDS program (run 3), and their Lick/IDS indices may be less
reliable. Therefore we do not use them here to calculate the offsets.
The fits are shown in Figure~\ref{fig_lickoffsets} and the offsets and
their 1$\sigma$ errors are listed in Table~\ref{tab_offsets}.
\begin{table}
  \begin{center}
    \caption{List of Lick/IDS indices measured in this work.}
    \label{tab_offsets}
    \begin{tabular}{lcr@{$\pm$}l}
      \hline
      index & $\sigma$ & \multicolumn{2}{c}{Offset} \\
            & (\kms)   & \multicolumn{2}{c}{(ours$-$Lick/IDS)} \\
      \hline
      CN$_1$ & 325 &$-$0.018 & 0.014 mag \\
      CN$_2$ & 325 &$-$0.023 & 0.022 mag \\
      \HdA   & 325 &   0.759 & 0.495 \AA \\
      \HdF   & 325 &   0.068 & 0.456 \AA \\
      Ca4227 & 300 &   0.065 & 0.289 \AA \\
      G4300  & 300 &   0.011 & 0.285 \AA \\
      \HgA   & 275 &$-$0.935 & 0.485 \AA \\
      \HgF   & 275 &$-$0.247 & 0.151 \AA \\
      Fe4383 & 250 &   0.097 & 0.622 \AA \\
      Ca4455 & 250 &$-$0.316 & 0.585 \AA \\
      Fe4531 & 250 &$-$0.380 & 0.365 \AA \\
      C4668  & 250 &   0.455 & 0.531 \AA \\
      \Hbeta & 225 &   0.130 & 0.174 \AA \\
      Fe5015 & 200 &   0.013 & 0.636 \AA \\
      Mg$_1$ & 200 &$-$0.021 & 0.006 mag \\
      Mg$_2$ & 200 &$-$0.018 & 0.010 mag \\
      Mg$b$  & 200 &   0.089 & 0.313 \AA \\
      Fe5270 & 200 &   0.196 & 0.191 \AA \\
      Fe5335 & 200 &   0.116 & 0.167 \AA \\
      Fe5406 & 200 &   0.146 & 0.119 \AA \\
      Fe5709 & 200 &$-$0.059 & 0.150 \AA \\
      Fe5782 & 200 &   0.047 & 0.103 \AA \\
      \hline
      \hline
    \end{tabular}
  \end{center}
\end{table}
\begin{table*}
  \caption{Comparison of indices measured for M\,32 in this work, in
    \citet{sanblas04thesis}, and in \citet{worthey04} (W04; indices at
    1.56\arcsec). \label{tab_compare_M32} }
  \begin{tabular}{lc*{3}{r@{ $\pm$ }l}}
    \hline
    index       & units & \multicolumn{2}{c}{this work} & \multicolumn{2}{c}{SB04} & \multicolumn{2}{c}{W04}\\
    \hline				   				  
    CN$_1$      & (mag) & $ 0.034$ & $0.002$ & $-0.020$ & $0.023$ & $ 0.022$ & $0.001$ \\
    CN$_2$      & (mag) & $ 0.062$ & $0.003$ & $ 0.013$ & $0.023$ & $ 0.053$ & $0.003$ \\
    H$\delta_A$ & (\AA) & $-1.342$ & $0.085$ & $-0.903$ & $0.110$ & $-1.043$ & $0.061$ \\
    H$\delta_F$ & (\AA) & $ 0.806$ & $0.057$ & $ 0.731$ & $0.038$ & $ 0.676$ & $0.045$ \\
    Ca4227      & (\AA) & $ 0.949$ & $0.043$ & $ 0.845$ & $0.032$ & $ 1.101$ & $0.025$ \\
    G4300       & (\AA) & $ 4.837$ & $0.073$ & $ 4.768$ & $0.111$ & $ 5.023$ & $0.050$ \\
    H$\gamma_A$ & (\AA) & $-3.851$ & $0.084$ & $-4.045$ & $0.572$ & $-4.308$ & $0.051$ \\
    H$\gamma_F$ & (\AA) & $-0.494$ & $0.051$ & $-0.155$ & $0.214$ & $-0.535$ & $0.021$ \\
    Fe4383      & (\AA) & $ 4.714$ & $0.104$ & $ 4.681$ & $0.386$ & $ 4.879$ & $0.073$ \\
    Ca4455      & (\AA) & $ 1.447$ & $0.055$ & $ 1.405$ & $0.117$ & $ 1.624$ & $0.032$ \\
    Fe4531      & (\AA) & $ 3.401$ & $0.082$ & $ 3.081$ & $0.154$ & $ 3.424$ & $0.060$ \\
    C4668       & (\AA) & $ 5.543$ & $0.125$ & $ 4.260$ & $0.831$ & $ 5.999$ & $0.142$ \\
    H$\beta$    & (\AA) & $ 1.977$ & $0.052$ & $ 2.214$ & $0.464$ & $ 2.190$ & $0.030$ \\
    Fe5015      & (\AA) & $ 5.079$ & $0.115$ & $ 5.220$ & $0.523$ & $ 5.219$ & $0.028$ \\
    Mg$_1$      & (mag) & $ 0.089$ & $0.001$ & \multicolumn{2}{c}{---} & $0.075$ & $0.001$ \\
    Mg$_2$      & (mag) & $ 0.207$ & $0.002$ & \multicolumn{2}{c}{---} & $0.198$ & $0.001$ \\
    Mg$b$       & (\AA) & $ 2.933$ & $0.058$ & $ 2.832$ & $0.328$ & $ 2.939$ & $0.048$ \\
    Fe5270      & (\AA) & $ 2.745$ & $0.064$ & $ 2.910$ & $0.102$ & $ 2.940$ & $0.020$ \\
    Fe5335      & (\AA) & $ 2.413$ & $0.073$ & $ 2.532$ & $0.069$ & $ 2.510$ & $0.032$ \\
    Fe5706      & (\AA) & $ 0.996$ & $0.055$ & \multicolumn{2}{c}{---} & $0.991$ & $0.038$ \\
    Fe5782      & (\AA) & $ 0.760$ & $0.047$ & \multicolumn{2}{c}{---} & $0.878$ & $0.021$ \\
    \hline
    \hline
  \end{tabular}
\end{table*}
We have one galaxy in common with the sample of SB06, namely M\,32.
In Table~\ref{tab_compare_M32}, we show the measurements of all the
indices in common for this galaxy. We also included measurements
obtained by \citet{worthey04}, at a radius of 1.56\arcsec (we summed
from -2\arcsec to 2\arcsec). They are in agreement within the error
bars.

\begin{figure*}
  \includegraphics[clip,width=8cm]{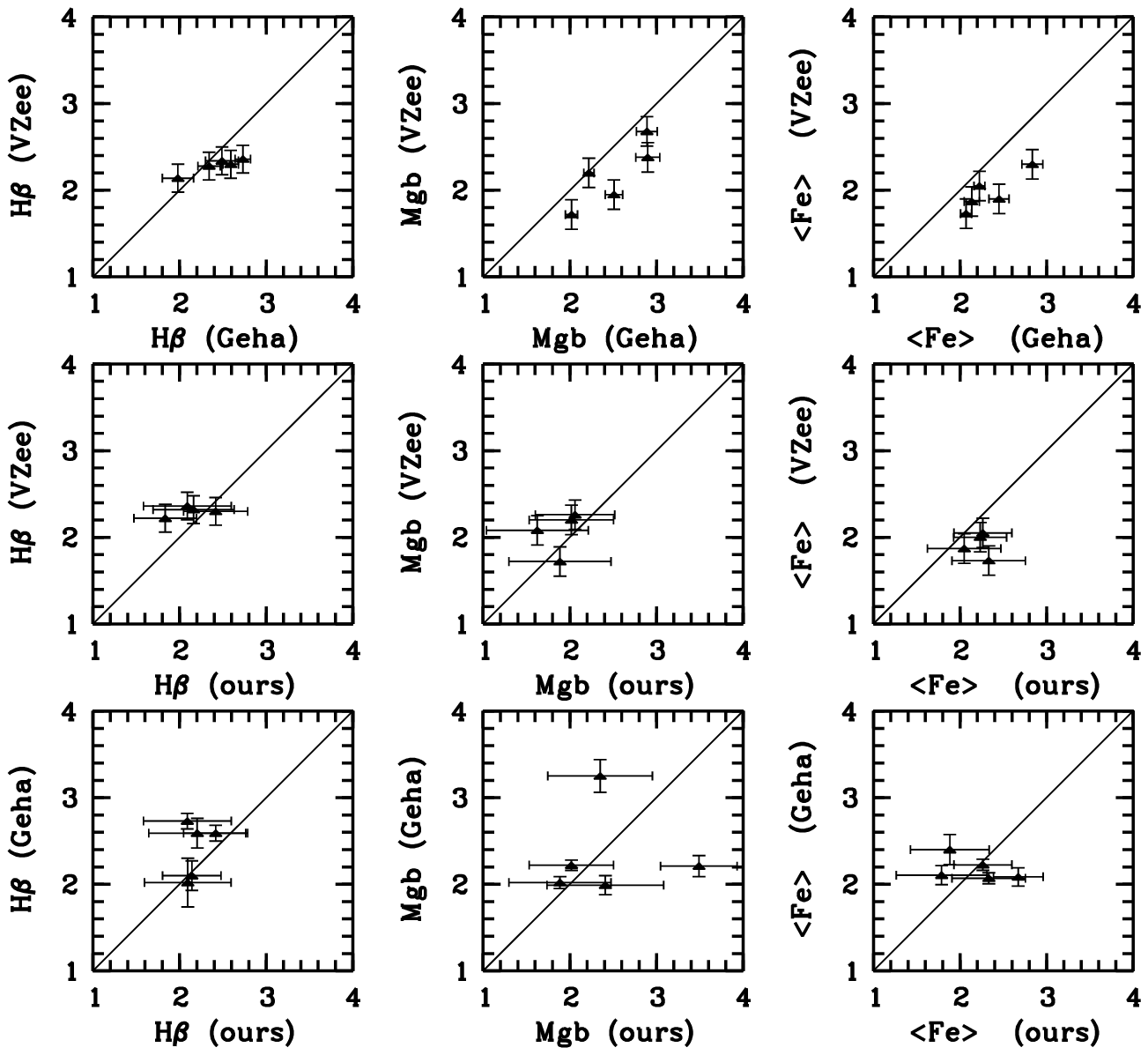}
  \caption{%
    Comparison between the Lick/IDS indices measured by us, \citet{gehaetal03}
    and \citet{vanzeeetal04}.}
  \label{fig_compare_our_geha_vzee}
\end{figure*}

We also have several galaxies in common with both \citet{gehaetal03}
and \citet{vanzeeetal04}. In Figure~\ref{fig_compare_our_geha_vzee},
we compare common measurements in those works and our present
work. Again the indices are in good agreement.

\section{Lick/IDS indices}
\label{app_indices}

In Table~\ref{tab_lickindices} we list, for all the galaxies, the
indices measured and their errors. The indices are transformed to the
Lick/IDS system using the offsets in Table~\ref{tab_offsets}.

\begin{table*}
  \begin{minipage}{\textwidth}
    \caption{Central line-strength indices corrected to the Lick/IDS
      system, measured in the central 4\arcsec. For each galaxy the
      first line are the indices and the second line are the 1$\sigma$ errors.}
    \label{tab_lickindices}
    \begin{tabular}{lcccccccccccc   }
\hline
galaxy & CN$_1$ & CN$_2$ & H$\delta_A$ & H$\delta_F$ & Ca4227 & G4300 & H$\gamma_A$ & H$\gamma_F$ & Fe4383 & Ca445 & Fe4531 & C4668 \\
       &(mag)&(mag)&(\AA)&(\AA)& (\AA) & (\AA) &(\AA)&(\AA)& (\AA) & (\AA) & (\AA) & (\AA) \\
\hline
M\,32     & $ 0.034$ & $ 0.062$ & $-1.342$ & $ 0.806$ & $ 0.949$ & $ 4.837$ & $-3.851$ & $-0.494$ & $ 4.714$ & $ 1.447$ & $ 3.401$ & $ 5.543$ \\
          & $ 0.002$ & $ 0.003$ & $ 0.085$ & $ 0.057$ & $ 0.043$ & $ 0.073$ & $ 0.084$ & $ 0.051$ & $ 0.104$ & $ 0.055$ & $ 0.082$ & $ 0.125$ \\
ID\,0650  & $-0.005$ & $ 0.023$ & $-0.588$ & $ 0.921$ & $ 1.159$ & $ 4.179$ & $-2.639$ & $ 0.270$ & $ 4.363$ & $ 1.172$ & $ 3.453$ & $ 4.246$ \\
          & $ 0.021$ & $ 0.025$ & $ 0.765$ & $ 0.521$ & $ 0.362$ & $ 0.648$ & $ 0.716$ & $ 0.436$ & $ 0.867$ & $ 0.472$ & $ 0.670$ & $ 1.004$ \\
ID\,0734  & $-0.070$ & $-0.042$ & $ 0.387$ & $ 1.985$ & $-0.105$ & $ 3.488$ & $-0.248$ & $ 1.664$ & $ 2.354$ & $ 1.242$ & $ 2.281$ & $ 0.568$ \\
          & $ 0.038$ & $ 0.045$ & $ 1.369$ & $ 0.936$ & $ 0.748$ & $ 1.225$ & $ 1.297$ & $ 0.814$ & $ 1.693$ & $ 0.873$ & $ 1.301$ & $ 1.955$ \\
ID\,0872  & $-0.100$ & $-0.065$ & $ 4.380$ & $ 3.416$ & $ 0.589$ & $ 0.667$ & $ 5.242$ & $ 4.111$ & $ 0.560$ & $ 0.602$ & $ 1.883$ & $-0.103$ \\
          & $ 0.037$ & $ 0.045$ & $ 1.240$ & $ 0.899$ & $ 0.683$ & $ 1.278$ & $ 1.191$ & $ 0.760$ & $ 1.819$ & $ 0.911$ & $ 1.398$ & $ 2.144$ \\
ID\,0918  & $ 0.020$ & $ 0.054$ & $-1.534$ & $ 0.721$ & $ 1.025$ & $ 5.165$ & $-3.537$ & $-0.416$ & $ 4.081$ & $ 1.433$ & $ 3.469$ & $ 3.911$ \\
          & $ 0.013$ & $ 0.015$ & $ 0.470$ & $ 0.315$ & $ 0.223$ & $ 0.388$ & $ 0.446$ & $ 0.266$ & $ 0.536$ & $ 0.276$ & $ 0.405$ & $ 0.616$ \\
ID\,1524  & $-0.073$ & $-0.032$ & $ 2.938$ & $ 2.660$ & $ 0.542$ & $ 2.037$ & $ 2.586$ & $ 3.228$ & $ 2.396$ & $ 0.881$ & $ 2.592$ & $ 1.110$ \\
          & $ 0.030$ & $ 0.037$ & $ 1.016$ & $ 0.726$ & $ 0.542$ & $ 1.028$ & $ 1.017$ & $ 0.613$ & $ 1.439$ & $ 0.753$ & $ 1.138$ & $ 1.747$ \\
VCC\,0021 & $-0.146$ & $-0.092$ & $ 7.251$ & $ 5.165$ & $ 0.248$ & $ 0.362$ & $ 6.628$ & $ 4.790$ & $ 0.521$ & $ 1.118$ & $ 1.341$ & $ 1.445$ \\
          & $ 0.021$ & $ 0.025$ & $ 0.635$ & $ 0.459$ & $ 0.366$ & $ 0.725$ & $ 0.640$ & $ 0.403$ & $ 1.025$ & $ 0.507$ & $ 0.829$ & $ 1.242$ \\
VCC\,0308 & $-0.054$ & $-0.003$ & $ 2.362$ & $ 2.528$ & $ 0.675$ & $ 3.231$ & $ 1.172$ & $ 2.439$ & $ 2.446$ & $ 0.885$ & $ 3.460$ & $ 3.427$ \\
          & $ 0.030$ & $ 0.036$ & $ 1.056$ & $ 0.747$ & $ 0.538$ & $ 0.943$ & $ 1.001$ & $ 0.627$ & $ 1.343$ & $ 0.737$ & $ 1.021$ & $ 1.529$ \\
VCC\,0397 & $-0.044$ & $-0.024$ & $ 1.383$ & $ 2.120$ & $ 0.737$ & $ 3.992$ & $ 0.062$ & $ 1.762$ & $ 2.594$ & $ 1.130$ & $ 3.365$ & $ 2.828$ \\
          & $ 0.027$ & $ 0.033$ & $ 0.966$ & $ 0.672$ & $ 0.479$ & $ 0.832$ & $ 0.898$ & $ 0.536$ & $ 1.188$ & $ 0.623$ & $ 0.883$ & $ 1.336$ \\
VCC\,0523 & $-0.020$ & $ 0.016$ & $ 1.045$ & $ 1.316$ & $ 1.085$ & $ 4.188$ & $-1.369$ & $ 0.689$ & $ 3.247$ & $ 0.825$ & $ 3.708$ & $ 3.875$ \\
          & $ 0.027$ & $ 0.032$ & $ 0.960$ & $ 0.689$ & $ 0.483$ & $ 0.824$ & $ 0.925$ & $ 0.570$ & $ 1.170$ & $ 0.627$ & $ 0.867$ & $ 1.344$ \\
VCC\,0856 & $-0.011$ & $ 0.027$ & $ 0.058$ & $ 1.111$ & $ 0.683$ & $ 3.796$ & $-2.260$ & $ 0.807$ & $ 4.607$ & $ 1.594$ & $ 2.573$ & $ 2.005$ \\
          & $ 0.025$ & $ 0.030$ & $ 0.884$ & $ 0.621$ & $ 0.460$ & $ 0.787$ & $ 0.867$ & $ 0.512$ & $ 1.082$ & $ 0.576$ & $ 0.873$ & $ 1.339$ \\
VCC\,0917 & $-0.038$ & $ 0.005$ & $ 0.155$ & $ 1.594$ & $ 0.669$ & $ 3.210$ & $-0.197$ & $ 0.842$ & $ 1.698$ & $ 1.534$ & $ 2.837$ & $ 0.766$ \\
          & $ 0.030$ & $ 0.035$ & $ 1.103$ & $ 0.764$ & $ 0.506$ & $ 0.878$ & $ 0.916$ & $ 0.572$ & $ 1.213$ & $ 0.610$ & $ 0.872$ & $ 1.290$ \\
VCC\,0990 & $-0.022$ & $ 0.011$ & $-0.094$ & $ 1.269$ & $ 0.938$ & $ 3.900$ & $-0.065$ & $ 0.782$ & $ 0.252$ & $ 1.233$ & $ 3.122$ & $ 2.019$ \\
          & $ 0.014$ & $ 0.017$ & $ 0.500$ & $ 0.345$ & $ 0.248$ & $ 0.446$ & $ 0.468$ & $ 0.298$ & $ 0.654$ & $ 0.335$ & $ 0.489$ & $ 0.759$ \\
VCC\,1087 & $-0.014$ & $ 0.011$ & $-0.540$ & $ 0.838$ & $ 0.908$ & $ 3.529$ & $-1.335$ & $ 0.968$ & $ 4.535$ & $ 1.387$ & $ 3.864$ & $ 2.969$ \\
          & $ 0.025$ & $ 0.030$ & $ 0.894$ & $ 0.613$ & $ 0.433$ & $ 0.782$ & $ 0.820$ & $ 0.493$ & $ 1.013$ & $ 0.522$ & $ 0.797$ & $ 1.167$ \\
VCC\,1122 & $-0.038$ & $-0.010$ & $ 0.512$ & $ 1.138$ & $ 0.634$ & $ 3.838$ & $-0.671$ & $ 1.333$ & $ 3.286$ & $ 1.404$ & $ 2.934$ & $ 1.900$ \\
          & $ 0.020$ & $ 0.024$ & $ 0.701$ & $ 0.492$ & $ 0.352$ & $ 0.630$ & $ 0.676$ & $ 0.409$ & $ 0.882$ & $ 0.453$ & $ 0.693$ & $ 1.062$ \\
VCC\,1183 & $-0.011$ & $ 0.023$ & $-0.367$ & $ 1.200$ & $ 1.013$ & $ 3.968$ & $-1.458$ & $ 0.753$ & $ 3.679$ & $ 1.053$ & $ 3.032$ & $ 3.232$ \\
          & $ 0.018$ & $ 0.021$ & $ 0.649$ & $ 0.442$ & $ 0.302$ & $ 0.538$ & $ 0.581$ & $ 0.356$ & $ 0.731$ & $ 0.389$ & $ 0.557$ & $ 0.833$ \\
VCC\,1261 & $-0.040$ & $-0.016$ & $ 0.440$ & $ 1.572$ & $ 1.054$ & $ 4.259$ & $-1.177$ & $ 0.982$ & $ 2.981$ & $ 0.971$ & $ 2.219$ & $ 3.429$ \\
          & $ 0.015$ & $ 0.019$ & $ 0.546$ & $ 0.380$ & $ 0.276$ & $ 0.486$ & $ 0.534$ & $ 0.326$ & $ 0.695$ & $ 0.368$ & $ 0.547$ & $ 0.807$ \\
VCC\,1431 & $ 0.028$ & $ 0.066$ & $-0.695$ & $ 0.786$ & $ 1.143$ & $ 4.314$ & $-2.219$ & $-0.185$ & $ 2.742$ & $ 1.515$ & $ 2.736$ & $ 2.500$ \\
          & $ 0.016$ & $ 0.019$ & $ 0.586$ & $ 0.403$ & $ 0.281$ & $ 0.493$ & $ 0.548$ & $ 0.342$ & $ 0.701$ & $ 0.366$ & $ 0.533$ & $ 0.814$ \\
VCC\,1549 & $ 0.033$ & $ 0.073$ & $-1.629$ & $ 0.789$ & $ 1.203$ & $ 5.104$ & $-3.617$ & $-0.565$ & $ 4.361$ & $ 1.621$ & $ 3.262$ & $ 3.385$ \\
          & $ 0.019$ & $ 0.023$ & $ 0.709$ & $ 0.481$ & $ 0.338$ & $ 0.593$ & $ 0.676$ & $ 0.410$ & $ 0.810$ & $ 0.422$ & $ 0.623$ & $ 0.950$ \\
VCC\,1695 & $-0.049$ & $ 0.000$ & $ 2.463$ & $ 2.767$ & $ 0.772$ & $ 3.364$ & $ 0.969$ & $ 2.074$ & $ 2.366$ & $ 1.194$ & $ 2.913$ & $ 2.101$ \\
          & $ 0.027$ & $ 0.032$ & $ 0.940$ & $ 0.660$ & $ 0.464$ & $ 0.822$ & $ 0.853$ & $ 0.524$ & $ 1.160$ & $ 0.613$ & $ 0.881$ & $ 1.323$ \\
VCC\,1861 & $-0.040$ & $-0.015$ & $-1.173$ & $ 1.354$ & $ 0.830$ & $ 4.514$ & $-2.554$ & $-0.365$ & $ 4.122$ & $ 1.352$ & $ 4.514$ & $ 2.928$ \\
          & $ 0.040$ & $ 0.048$ & $ 1.505$ & $ 1.012$ & $ 0.718$ & $ 1.224$ & $ 1.364$ & $ 0.824$ & $ 1.632$ & $ 0.847$ & $ 1.194$ & $ 1.841$ \\
VCC\,1910 & $ 0.031$ & $ 0.050$ & $-1.741$ & $ 0.373$ & $ 1.152$ & $ 4.858$ & $-2.977$ & $-0.392$ & $ 4.338$ & $ 1.619$ & $ 4.067$ & $ 5.319$ \\
          & $ 0.018$ & $ 0.021$ & $ 0.656$ & $ 0.449$ & $ 0.304$ & $ 0.530$ & $ 0.591$ & $ 0.362$ & $ 0.718$ & $ 0.367$ & $ 0.533$ & $ 0.805$ \\
VCC\,1912 & $-0.077$ & $-0.036$ & $ 3.720$ & $ 3.212$ & $ 0.462$ & $ 1.754$ & $ 3.535$ & $ 3.222$ & $ 1.294$ & $ 1.038$ & $ 2.938$ & $ 2.044$ \\
          & $ 0.010$ & $ 0.012$ & $ 0.327$ & $ 0.232$ & $ 0.182$ & $ 0.341$ & $ 0.326$ & $ 0.201$ & $ 0.498$ & $ 0.252$ & $ 0.386$ & $ 0.604$ \\
VCC\,1947 & $-0.007$ & $ 0.022$ & $-1.352$ & $ 0.621$ & $ 1.194$ & $ 4.937$ & $-4.070$ & $-0.879$ & $ 4.654$ & $ 1.618$ & $ 3.655$ & $ 4.763$ \\
          & $ 0.015$ & $ 0.018$ & $ 0.579$ & $ 0.399$ & $ 0.272$ & $ 0.459$ & $ 0.535$ & $ 0.322$ & $ 0.627$ & $ 0.327$ & $ 0.477$ & $ 0.717$ \\
\hline
\hline
    \end{tabular}
  \end{minipage}
\end{table*}

\begin{table*}
  \begin{minipage}{\textwidth}
    \addtocounter{table}{-1}
    \caption{continued}
    \begin{tabular}{lcccccccccccc}
\hline
galaxy & H$\beta$ & Fe5015 & Mg$_1$ & Mg$_2$ & Mg$b$ & Fe5270 & Fe5335 & Fe5406 & Fe5709 & Fe5782 & D4000 \\
    & (\AA) & (\AA) &(mag)&(mag)&(\AA)& (\AA) & (\AA) & (\AA) & (\AA) & (\AA) & (\AA) \\
\hline
M\,32     & $ 1.977$ & $ 5.079$ & $ 0.089$ & $ 0.207$ & $ 2.933$ & $ 2.745$ & $ 2.413$ & $ 1.504$ & $ 0.996$ & $ 0.760$ & $2.0424$ \\
          & $ 0.052$ & $ 0.115$ & $ 0.001$ & $ 0.002$ & $ 0.058$ & $ 0.064$ & $ 0.073$ & $ 0.055$ & $ 0.047$ & $ 0.044$ & $0.0029$ \\
ID\,0650  & $ 1.992$ & $ 4.860$ & $ 0.054$ & $ 0.157$ & $ 2.284$ & $ 2.788$ & $ 2.273$ & $ 1.420$ & $ 0.849$ & $ 0.540$ & $1.9015$ \\
          & $ 0.418$ & $ 0.884$ & $ 0.010$ & $ 0.012$ & $ 0.461$ & $ 0.499$ & $ 0.566$ & $ 0.430$ & $ 0.375$ & $ 0.360$ & $0.0277$ \\
ID\,0734  & $ 2.205$ & $ 2.939$ & $ 0.015$ & $ 0.083$ & $ 1.806$ & $ 2.186$ & $ 2.097$ & $ 0.473$ & $ 0.428$ & $-0.204$ & $1.6989$ \\
          & $ 0.751$ & $ 1.647$ & $ 0.018$ & $ 0.021$ & $ 0.831$ & $ 0.913$ & $ 1.063$ & $ 0.832$ & $ 0.745$ & $ 0.737$ & $0.0438$ \\
ID\,0872  & $ 3.249$ & $ 1.788$ & $ 0.028$ & $ 0.089$ & $ 1.264$ & $ 1.562$ & $ 1.489$ & $ 0.816$ & $ 1.286$ & $ 0.493$ & $1.5322$ \\
          & $ 0.847$ & $ 1.911$ & $ 0.020$ & $ 0.024$ & $ 0.973$ & $ 1.090$ & $ 1.264$ & $ 0.945$ & $ 0.829$ & $ 0.833$ & $0.0358$ \\
ID\,0918  & $ 2.121$ & $ 4.687$ & $ 0.080$ & $ 0.201$ & $ 2.944$ & $ 2.567$ & $ 2.048$ & $ 1.419$ & $ 0.952$ & $ 0.531$ & $2.1667$ \\
          & $ 0.249$ & $ 0.537$ & $ 0.006$ & $ 0.007$ & $ 0.269$ & $ 0.294$ & $ 0.335$ & $ 0.252$ & $ 0.209$ & $ 0.202$ & $0.0196$ \\
ID\,1524  & $ 2.838$ & $ 2.804$ & $ 0.031$ & $ 0.094$ & $ 1.315$ & $ 1.895$ & $ 0.751$ & $ 0.805$ & $ 0.642$ & $ 0.607$ & $1.5575$ \\
          & $ 0.714$ & $ 1.582$ & $ 0.017$ & $ 0.022$ & $ 0.877$ & $ 0.917$ & $ 1.116$ & $ 0.819$ & $ 0.737$ & $ 0.730$ & $0.0293$ \\
VCC\,0021 & $ 4.172$ & $ 2.608$ & $ 0.022$ & $ 0.060$ & $ 1.094$ & $ 1.086$ & $ 1.492$ & $ 0.250$ & $ 0.572$ & $ 0.294$ & $1.5113$ \\
          & $ 0.486$ & $ 1.102$ & $ 0.012$ & $ 0.014$ & $ 0.560$ & $ 0.649$ & $ 0.730$ & $ 0.565$ & $ 0.480$ & $ 0.474$ & $0.0174$ \\
VCC\,0308 & $ 2.637$ & $ 3.520$ & $ 0.026$ & $ 0.102$ & $ 2.209$ & $ 2.121$ & $ 1.928$ & $ 0.547$ & $ 0.831$ & $ 0.412$ & $1.6870$ \\
          & $ 0.610$ & $ 1.324$ & $ 0.015$ & $ 0.018$ & $ 0.675$ & $ 0.747$ & $ 0.886$ & $ 0.688$ & $ 0.591$ & $ 0.583$ & $0.0319$ \\
VCC\,0397 & $ 2.850$ & $ 4.607$ & $ 0.041$ & $ 0.136$ & $ 2.209$ & $ 2.698$ & $ 1.831$ & $ 1.458$ & $ 1.100$ & $ 0.551$ & $1.7788$ \\
          & $ 0.529$ & $ 1.128$ & $ 0.012$ & $ 0.015$ & $ 0.580$ & $ 0.634$ & $ 0.712$ & $ 0.546$ & $ 0.453$ & $ 0.434$ & $0.0306$ \\
VCC\,0523 & $ 2.423$ & $ 3.926$ & $ 0.053$ & $ 0.144$ & $ 2.500$ & $ 2.419$ & $ 1.567$ & $ 1.111$ & $ 1.110$ & $ 0.412$ & $1.7982$ \\
          & $ 0.552$ & $ 1.186$ & $ 0.013$ & $ 0.015$ & $ 0.591$ & $ 0.652$ & $ 0.741$ & $ 0.552$ & $ 0.447$ & $ 0.435$ & $0.0320$ \\
VCC\,0856 & $ 2.203$ & $ 2.845$ & $ 0.046$ & $ 0.134$ & $ 2.410$ & $ 2.014$ & $ 1.560$ & $ 0.882$ & $ 0.938$ & $ 0.671$ & $1.7672$ \\
          & $ 0.532$ & $ 1.190$ & $ 0.013$ & $ 0.015$ & $ 0.597$ & $ 0.666$ & $ 0.766$ & $ 0.582$ & $ 0.496$ & $ 0.479$ & $0.0301$ \\
VCC\,0917 & $ 2.089$ & $ 3.633$ & $ 0.033$ & $ 0.121$ & $ 1.886$ & $ 2.365$ & $ 2.300$ & $ 1.161$ & $ 0.794$ & $ 0.103$ & $1.9108$ \\
          & $ 0.476$ & $ 1.004$ & $ 0.011$ & $ 0.013$ & $ 0.500$ & $ 0.540$ & $ 0.603$ & $ 0.459$ & $ 0.387$ & $ 0.376$ & $0.0386$ \\
VCC\,0990 & $ 1.834$ & $ 4.454$ & $ 0.048$ & $ 0.140$ & $ 2.060$ & $ 2.420$ & $ 2.045$ & $ 1.061$ & $ 0.535$ & $ 0.638$ & $1.7946$ \\
          & $ 0.316$ & $ 0.651$ & $ 0.007$ & $ 0.009$ & $ 0.336$ & $ 0.368$ & $ 0.417$ & $ 0.320$ & $ 0.280$ & $ 0.256$ & $0.0156$ \\
VCC\,1087 & $ 2.093$ & $ 4.531$ & $ 0.069$ & $ 0.165$ & $ 2.350$ & $ 2.022$ & $ 1.742$ & $ 1.861$ & $ 0.713$ & $ 0.483$ & $1.9120$ \\
          & $ 0.469$ & $ 0.998$ & $ 0.011$ & $ 0.014$ & $ 0.518$ & $ 0.566$ & $ 0.662$ & $ 0.485$ & $ 0.445$ & $ 0.414$ & $0.0348$ \\
VCC\,1122 & $ 2.161$ & $ 4.620$ & $ 0.034$ & $ 0.129$ & $ 1.625$ & $ 2.055$ & $ 2.041$ & $ 0.840$ & $ 0.333$ & $ 0.710$ & $1.7931$ \\
          & $ 0.434$ & $ 0.924$ & $ 0.010$ & $ 0.012$ & $ 0.497$ & $ 0.538$ & $ 0.602$ & $ 0.462$ & $ 0.404$ & $ 0.380$ & $0.0235$ \\
VCC\,1183 & $ 2.416$ & $ 4.685$ & $ 0.055$ & $ 0.144$ & $ 2.186$ & $ 2.560$ & $ 1.879$ & $ 1.173$ & $ 1.098$ & $ 0.546$ & $2.0361$ \\
          & $ 0.333$ & $ 0.724$ & $ 0.008$ & $ 0.010$ & $ 0.369$ & $ 0.408$ & $ 0.456$ & $ 0.344$ & $ 0.295$ & $ 0.286$ & $0.0252$ \\
VCC\,1261 & $ 2.416$ & $ 3.947$ & $ 0.040$ & $ 0.144$ & $ 2.018$ & $ 2.216$ & $ 2.310$ & $ 1.397$ & $ 0.703$ & $ 0.337$ & $1.8308$ \\
          & $ 0.328$ & $ 0.725$ & $ 0.008$ & $ 0.010$ & $ 0.374$ & $ 0.415$ & $ 0.460$ & $ 0.351$ & $ 0.308$ & $ 0.296$ & $0.0188$ \\
VCC\,1431 & $ 1.599$ & $ 3.645$ & $ 0.074$ & $ 0.178$ & $ 2.829$ & $ 2.140$ & $ 1.644$ & $ 1.023$ & $ 0.729$ & $ 0.221$ & $1.8292$ \\
          & $ 0.320$ & $ 0.698$ & $ 0.007$ & $ 0.009$ & $ 0.344$ & $ 0.380$ & $ 0.437$ & $ 0.329$ & $ 0.277$ & $ 0.268$ & $0.0188$ \\
VCC\,1549 & $ 1.710$ & $ 4.420$ & $ 0.077$ & $ 0.200$ & $ 3.277$ & $ 2.099$ & $ 1.976$ & $ 1.308$ & $ 1.094$ & $ 0.637$ & $2.0037$ \\
          & $ 0.396$ & $ 0.843$ & $ 0.009$ & $ 0.011$ & $ 0.425$ & $ 0.473$ & $ 0.531$ & $ 0.398$ & $ 0.343$ & $ 0.329$ & $0.0265$ \\
VCC\,1695 & $ 2.880$ & $ 4.447$ & $ 0.032$ & $ 0.107$ & $ 1.797$ & $ 1.573$ & $ 1.899$ & $ 0.986$ & $ 0.997$ & $ 0.646$ & $1.7142$ \\
          & $ 0.509$ & $ 1.095$ & $ 0.012$ & $ 0.014$ & $ 0.560$ & $ 0.633$ & $ 0.704$ & $ 0.534$ & $ 0.444$ & $ 0.424$ & $0.0283$ \\
VCC\,1861 & $ 1.931$ & $ 4.246$ & $ 0.045$ & $ 0.130$ & $ 2.911$ & $ 2.070$ & $ 1.894$ & $ 1.380$ & $ 1.093$ & $ 0.530$ & $1.9882$ \\
          & $ 0.742$ & $ 1.583$ & $ 0.017$ & $ 0.021$ & $ 0.779$ & $ 0.885$ & $ 0.995$ & $ 0.755$ & $ 0.688$ & $ 0.671$ & $0.0603$ \\
VCC\,1910 & $ 1.886$ & $ 4.809$ & $ 0.087$ & $ 0.206$ & $ 3.157$ & $ 2.678$ & $ 2.282$ & $ 1.643$ & $ 0.745$ & $ 0.668$ & $2.1377$ \\
          & $ 0.335$ & $ 0.719$ & $ 0.008$ & $ 0.010$ & $ 0.363$ & $ 0.398$ & $ 0.448$ & $ 0.337$ & $ 0.294$ & $ 0.281$ & $0.0282$ \\
VCC\,1912 & $ 3.705$ & $ 3.281$ & $ 0.030$ & $ 0.104$ & $ 1.329$ & $ 1.595$ & $ 1.490$ & $ 0.739$ & $ 0.783$ & $ 0.377$ & $1.5998$ \\
          & $ 0.238$ & $ 0.552$ & $ 0.006$ & $ 0.007$ & $ 0.284$ & $ 0.319$ & $ 0.365$ & $ 0.276$ & $ 0.237$ & $ 0.227$ & $0.0094$ \\
VCC\,1947 & $ 2.141$ & $ 5.708$ & $ 0.087$ & $ 0.222$ & $ 3.491$ & $ 3.035$ & $ 2.310$ & $ 1.800$ & $ 1.190$ & $ 0.561$ & $2.0996$ \\
          & $ 0.289$ & $ 0.623$ & $ 0.007$ & $ 0.008$ & $ 0.312$ & $ 0.343$ & $ 0.386$ & $ 0.288$ & $ 0.241$ & $ 0.235$ & $0.0228$ \\
\hline
\hline
    \end{tabular}
  \end{minipage}
\end{table*}

\label{lastpage}

\begin{thebibliography}{99}
  
\bibitem[\protect\citeauthoryear{Beasley et al.}{2006}]{beasleyetal06} Beasley
  M.~A., Strader J., Brodie J.~P., Cenarro A.~J., Geha M., 2006, AJ, 131, 814
  
\bibitem[\protect\citeauthoryear{Binggeli, Sandage \&
    Tammann}{1985}]{binggelietal85} Binggeli B., Sandage A., Tammann G.~A.,
  1985, AJ, 90, 1681
  
\bibitem[\protect\citeauthoryear{Binggeli, Tammann \&
    Sandage}{1987}]{binggelietal87} Binggeli B., Tammann G.~A., Sandage A.,
  1987, AJ, 94, 251
  
\bibitem[\protect\citeauthoryear{Boselli et al.}{2007}]{boselli07} Boselli A.,
  Boissier S., Cortese L. \& Gavazzi G., 2007, ApJ, submitted
  
\bibitem[\protect\citeauthoryear{Boselli \& Gavazzi}{2006}]{boselligavazzi06}
  Boselli A., Gavazzi G., 2006, PASP, 118, 517
  
\bibitem[\protect\citeauthoryear{Boselli et al.}{2001}]{bosellietal01} Boselli
  A., Gavazzi G., Donas J., Scodeggio M., 2001, AJ, 121, 753
  
\bibitem[\protect\citeauthoryear{Bouchard et al.}{2005}]{bouchardetal05}
  Bouchard A., Jerjen H., Da Costa G.~S., Ott J., 2005, AJ, 130, 2058
  
\bibitem[\protect\citeauthoryear{Bundy et al.}{2006}]{bundyetal06} Bundy K.,
  et al., 2006, ApJ, 651, 120
  
\bibitem[\protect\citeauthoryear{Burstein et
al.}{1984}]{bursteinetal84} Burstein D., Faber S.~M., Gaskell C.~M.,
Krumm N., 1984, ApJ, 287, 586

\bibitem[\protect\citeauthoryear{Burstein et
  al.}{2004}]{bursteinetal04} Burstein D., et al., 2004, ApJ, 614, 158

\bibitem[\protect\citeauthoryear{Buyle et al.}{2005}]{buyleetal05} Buyle P.,
  De Rijcke S., Michielsen D., Baes M., Dejonghe H., 2005, MNRAS, 360, 853
  
\bibitem[\protect\citeauthoryear{Caldwell, Rose \&
    Concannon}{2003}]{caldwelletal03} Caldwell N., Rose J.~A., Concannon
  K.~D., 2003, AJ, 125, 2891
  
\bibitem[\protect\citeauthoryear{Cardiel}{1999}]{cardiel99} Cardiel N., 1999,
  Ph. D. Thesis, Universidad Complutense de Madrid
  
\bibitem[\protect\citeauthoryear{Cardiel et al.}{2003}]{cardieletal03} Cardiel
  N., Gorgas J., S{\'a}nchez-Bl{\'a}zquez P., Cenarro A.~J., Pedraz S.,
  Bruzual G., Klement J., 2003, A\&A, 409, 511
  
\bibitem[\protect\citeauthoryear{Carraro et al.}{2001}]{carraroetal01} Carraro
  G., Chiosi C., Girardi L., Lia C., 2001, MNRAS, 327, 69
  
\bibitem[\protect\citeauthoryear{Cenarro et al.}{2003}]{cenarroetal03} Cenarro
  A.~J., Gorgas J., Vazdekis A., Cardiel N., Peletier R.~F., 2003, MNRAS, 339,
  L12
  
\bibitem[\protect\citeauthoryear{Cenarro et al.}{2004}]{cenarroetal04} Cenarro
  A.~J., S{\'a}nchez-Bl{\'a}zquez P., Cardiel N., Gorgas J., 2004, ApJ, 614,
  L101
  
\bibitem[\protect\citeauthoryear{Cenarro et al.}{2007}]{cenarroetal07} Cenarro
  A.~J., Beasley M.~A., Strader J., Brodie J.~P., Forbes D.~A., 2007, AJ, 134,
  391
  
\bibitem[\protect\citeauthoryear{Conselice}{2003}]{conselice03} Conselice
  C.~J., 2003, ApJS, 147, 1
  
\bibitem[\protect\citeauthoryear{Conselice}{2006}]{conselice06} Conselice
  C.~J., 2006, ApJ, 639, 120

\bibitem[\protect\citeauthoryear{Conselice, Bershady, \&
Jangren}{2000}]{conselice00} Conselice C.~J., Bershady M.~A., Jangren
A., 2000, ApJ, 529, 886

\bibitem[\protect\citeauthoryear{Conselice, Gallagher \&
    Wyse}{2001}]{conseliceetal01} Conselice C.~J., Gallagher J.~S., III, Wyse
  R.~F.~G., 2001, ApJ, 559, 791
  
\bibitem[\protect\citeauthoryear{Conselice et al.}{2003b}]{conseliceetal03HI}
  Conselice C.~J., O'Neil K., Gallagher J.~S., Wyse R.~F.~G., 2003, ApJ, 591,
  167
  
\bibitem[\protect\citeauthoryear{Cowie et al.}{1996}]{cowieetal96} Cowie
  L.~L., Songaila A., Hu E.~M., Cohen J.~G., 1996, AJ, 112, 839
  
\bibitem[\protect\citeauthoryear{Davies \& Phillipps}{1988}]{daviesphillips88}
  Davies J.~I., Phillipps S., 1988, MNRAS, 233, 553
  
\bibitem[\protect\citeauthoryear{Dekel \& Woo}{2003}]{dekelwoo03} Dekel A.,
  Woo J., 2003, MNRAS, 344, 1131
  
\bibitem[\protect\citeauthoryear{de la Rosa et al.}{2007}]{delarosa07}
de la Rosa I.~G., de Carvalho R.~R., Vazdekis A., Barbuy B., 2007, AJ,
133, 330

\bibitem[\protect\citeauthoryear{De Lucia et al.}{2006}]{deluciaetal06} De
  Lucia G., Springel V., White S.~D.~M., Croton D., Kauffmann G., 2006, MNRAS,
  366, 499
  
\bibitem[\protect\citeauthoryear{De Rijcke et
    al.}{2003a}]{derijckeetal03disks} De Rijcke S., Dejonghe H., Zeilinger
  W.~W., Hau G.~K.~T., 2003, A\&A, 400, 119
  
\bibitem[\protect\citeauthoryear{de Rijcke et al.}{2005}]{derijckeetal05} De
  Rijcke S., Michielsen D., Dejonghe H., Zeilinger W.~W., Hau G.~K.~T., 2005,
  A\&A, 438, 491
  
\bibitem[\protect\citeauthoryear{De Rijcke et al.}{2003b}]{derijckeetal03ism}
  De Rijcke S., Zeilinger W.~W., Dejonghe H., Hau G.~K.~T., 2003, MNRAS, 339,
  225
  
\bibitem[\protect\citeauthoryear{Ferguson \&
    Binggeli}{1994}]{fergusonbinggeli94} Ferguson H.~C., Binggeli B., 1994,
  A\&ARv, 6, 67
  
\bibitem[\protect\citeauthoryear{Ganda et al.}{2007}]{gandaetal07} Ganda K. et
  al., 2007, MNRAS, accepted (arXiv:0706.3624)
  
\bibitem[\protect\citeauthoryear{Gavazzi et al.}{2003}]{gavazzi03goldmine}
  Gavazzi G., Boselli A., Donati A., Franzetti P., Scodeggio M., 2003, A\&A,
  400, 451
  
\bibitem[\protect\citeauthoryear{Gavazzi et al.}{1999}]{gavazzietal99} Gavazzi
  G., Boselli A., Scodeggio M., Pierini D., Belsole E., 1999, MNRAS, 304, 595
  
\bibitem[\protect\citeauthoryear{Gavazzi, Pierini \&
    Boselli}{1996}]{gavazzietal96} Gavazzi G., Pierini D., Boselli A., 1996,
  A\&A, 312, 397
  
\bibitem[\protect\citeauthoryear{Gavazzi et al.}{2001}]{gavazzietal01} Gavazzi
  G., Zibetti S., Boselli A., Franzetti P., Scodeggio M., Martocchi S., 2001,
  A\&A, 372, 29
  
\bibitem[\protect\citeauthoryear{Geha, Guhathakurta \& van der
    Marel}{2003}]{gehaetal03} Geha M., Guhathakurta P., van der Marel R.~P.,
  2003, AJ, 126, 1794
  
\bibitem[\protect\citeauthoryear{Gorgas, Jablonka \&
    Goudfrooij}{2007}]{gorgasetal07} Gorgas J., Jablonka P., Goudfrooij P.,
  2007, A\&A, submitted
  
\bibitem[\protect\citeauthoryear{Gorgas et al.}{1997}]{gorgasetal97} Gorgas
  J., Pedraz S., Guzman R., Cardiel N., Gonzalez J.~J., 1997, ApJ, 481, L19
  
\bibitem[\protect\citeauthoryear{Grebel, Gallagher \&
    Harbeck}{2003}]{grebeletal03} Grebel E.~K., Gallagher J.~S., III, Harbeck
  D., 2003, AJ, 125, 1926
  
\bibitem[\protect\citeauthoryear{Gunn \& Gott}{1972}]{gunngott72} Gunn J.~E.,
  Gott J.~R.~I., 1972, ApJ, 176, 1
  
\bibitem[\protect\citeauthoryear{Li \& Burstein}{2003}]{liburstein03} Li Y.,
  Burstein D., 2003, ApJ, 598, L103
  
\bibitem[\protect\citeauthoryear{Lisker et al.}{2006}]{liskeretal06bluenuc}
  Lisker T., Glatt K., Westera P., Grebel E.~K., 2006, AJ, 132, 2432
  
\bibitem[\protect\citeauthoryear{Lisker, Grebel \&
    Binggeli}{2006}]{liskeretal06disk} Lisker T., Grebel E.~K., Binggeli B.,
  2006, AJ, 132, 497
  
\bibitem[\protect\citeauthoryear{Lisker et al.}{2007}]{liskeretal07} Lisker
  T., Grebel E.~K., Binggeli B., Glatt K., 2007, ApJ, 660, 1186
  
\bibitem[\protect\citeauthoryear{Mac Low \& Ferrara}{1999}]{maclowferrara99}
  Mac Low M.-M., Ferrara A., 1999, ApJ, 513, 142
  
\bibitem[\protect\citeauthoryear{Maraston et al.}{2003}]{maraston03}
Maraston C., Greggio L., Renzini A., Ortolani S., Saglia R.~P., Puzia
T.~H., Kissler-Patig M., 2003, A\&A, 400, 823

\bibitem[\protect\citeauthoryear{Marcolini, Brighenti \&
    D'Ercole}{2003}]{marcolinietal03} Marcolini A., Brighenti F., D'Ercole A.,
  2003, MNRAS, 345, 1329
  
\bibitem[\protect\citeauthoryear{Mateo}{1998}]{mateo98} Mateo M.~L., 1998,
  ARA\&A, 36, 435
  
\bibitem[\protect\citeauthoryear{Meynet \& Maeder}{2002}]{meynetmaeder02}
  Meynet G., Maeder A., 2002, A\&A, 390, 561
  
\bibitem[\protect\citeauthoryear{Michielsen et al.}{2004}]{michielsenetal04}
  Michielsen D., De Rijcke S., Zeilinger W.~W., Prugniel P., Dejonghe H.,
  Roberts S., 2004, MNRAS, 353, 1293
  
\bibitem[\protect\citeauthoryear{Michielsen et al.}{2003}]{michielsenetal03}
  Michielsen D., De Rijcke S., Dejonghe H., Zeilinger W.~W., Hau G.~K.~T.,
  2003, ApJ, 597, L21
  
\bibitem[\protect\citeauthoryear{Moore, Lake \& Katz}{1998}]{mooreetal98}
  Moore B., Lake G., Katz N., 1998, ApJ, 495, 139
  
\bibitem[\protect\citeauthoryear{Mori \& Burkert}{2000}]{moriburkert00} Mori
  M., Burkert A., 2000, ApJ, 538, 559
  
\bibitem[\protect\citeauthoryear{Nelan et al.}{2005}]{nelanetal05} Nelan
  J.~E., Smith R.~J., Hudson M.~J., Wegner G.~A., Lucey J.~R., Moore S.~A.~W.,
  Quinney S.~J., Suntzeff N.~B., 2005, ApJ, 632, 137
  
\bibitem[\protect\citeauthoryear{Paturel et al.}{2003}]{paturel03hyperleda}
  Paturel G., Petit C., Prugniel P., Theureau G., Rousseau J., Brouty M.,
  Dubois P., Cambr{\'e}sy L., 2003, A\&A, 412, 45

  
\bibitem[\protect\citeauthoryear{Pedraz et al.}{2002}]{pedrazetal02} Pedraz
  S., Gorgas J., Cardiel N., S{\'a}nchez-Bl{\'a}zquez P., Guzm{\'a}n R., 2002,
  MNRAS, 332, L59

\bibitem[\protect\citeauthoryear{Peletier et al.}{2007}]{peletier07}
  Peletier R.~F., et al., 2007, in preparation
  
\bibitem[\protect\citeauthoryear{Poggianti et al.}{2001}]{poggiantietal01}
  Poggianti B.~M., et al., 2001, ApJ, 562, 689
  
\bibitem[\protect\citeauthoryear{Prochaska, Rose \&
    Schiavon}{2005}]{prochaskaetal05} Prochaska L.~C., Rose J.~A., Schiavon
  R.~P., 2005, AJ, 130, 2666
  
\bibitem[\protect\citeauthoryear{Simien \&
  Prugniel}{2002}]{simienprugniel02} Simien F., Prugniel P., 2002,
  A\&A, 384, 371
  
\bibitem[\protect\citeauthoryear{S{\'a}nchez-Bl{\'a}zquez}{2007}]{sanblas04thesis}
  S{\'a}nchez-Bl{\'a}zquez P., 2004, Ph. D. Thesis, Universidad
  Complutense de Madrid

\bibitem[\protect\citeauthoryear{S{\'a}nchez-Bl{\'a}zquez et
    al.}{2006a}]{sanblas06} S{\'a}nchez-Bl{\'a}zquez P., Gorgas J.,
    Cardiel N., Gonz{\'a}lez J.~J., 2006a, A\&A, 457, 787 (SB06)
  
\bibitem[\protect\citeauthoryear{S{\'a}nchez-Bl{\'a}zquez et
    al.}{2006b}]{sanblasetal06b} S{\'a}nchez-Bl{\'a}zquez P., Gorgas
    J., Cardiel N., Gonz{\'a}lez J.~J., 2006b, A\&A, 457, 809
  
\bibitem[\protect\citeauthoryear{S{\'a}nchez-Bl{\'a}zquez et
    al.}{2006c}]{sanblas06miles} S{\'a}nchez-Bl{\'a}zquez P., et al.,
    2006c, MNRAS, 371, 703
  
\bibitem[\protect\citeauthoryear{Sandage, Binggeli \&
    Tammann}{1985}]{sandageetal85} Sandage A., Binggeli B., Tammann G.~A.,
  1985, AJ, 90, 1759
  
\bibitem[\protect\citeauthoryear{Rakos et al.}{2001}]{rakosetal01} Rakos K.,
  Schombert J., Maitzen H.~M., Prugovecki S., Odell A., 2001, AJ, 121, 1974
  
\bibitem[\protect\citeauthoryear{Smith et al.}{2006}]{smithetal06} Smith
  R.~J., Hudson M.~J., Lucey J.~R., Nelan J.~E., Wegner G.~A., 2006, MNRAS,
  369, 1419
 
\bibitem[\protect\citeauthoryear{Thomas, Maraston \& Bender}{2003a}]{TMB03}
  Thomas D., Maraston C., Bender R., 2003, MNRAS, 339, 897 (TMB03)
  
\bibitem[\protect\citeauthoryear{Thomas, Maraston \&
    Bender}{2003b}]{thomasetal03} Thomas D., Maraston C., Bender R., 2003,
  MNRAS, 343, 279
  
\bibitem[\protect\citeauthoryear{Thomas et al.}{2005}]{thomasetal05} Thomas
  D., Maraston C., Bender R., Mendes de Oliveira C., 2005, ApJ, 621, 673
 
\bibitem[\protect\citeauthoryear{Thomas, Maraston \&
    Korn}{2004}]{thomasetal04} Thomas D., Maraston C., Korn A., 2004, MNRAS,
  351, L19
  
\bibitem[\protect\citeauthoryear{Toloba et al.}{2007}]{toloba07} Toloba E.  et
  al., 2007, in preparation
  
\bibitem[\protect\citeauthoryear{Trager et al.}{2000}]{trageretal00} Trager
  S.~C., Faber S.~M., Worthey G., Gonz{\'a}lez J.~J., 2000, AJ, 120, 165

\bibitem[\protect\citeauthoryear{Trentham \& Tully}{2002}]{threnthamtully02}
  Trentham N., Tully R.~B., 2002, MNRAS, 335, 712
  
\bibitem[\protect\citeauthoryear{Tripicco \& Bell}{1995}]{tripiccobell95}
  Tripicco M.~J., Bell R.~A., 1995, AJ, 110, 3035
  
\bibitem[\protect\citeauthoryear{Vader}{1986}]{vader86} Vader J.~P., 1986,
  ApJ, 305, 669
  
\bibitem[\protect\citeauthoryear{van Zee, Barton \&
    Skillman}{2004}]{vanzeeetal04} van Zee L., Barton E.~J., Skillman E.~D.,
  2004, AJ, 128, 2797
  
\bibitem[\protect\citeauthoryear{van Zee, Skillman \&
    Haynes}{2004}]{vanzeeetal04rotation} van Zee L., Skillman E.~D., Haynes
  M.~P., 2004, AJ, 128, 121
  
\bibitem[\protect\citeauthoryear{Vazdekis}{1999}]{vazdekis99} Vazdekis A.,
  1999, ApJ, 513, 224
  
\bibitem[\protect\citeauthoryear{Vazdekis et al.}{1996}]{vazdekis96} Vazdekis
  A., Casuso E., Peletier R.~F., Beckman J.~E., 1996, ApJS, 106, 307 (V96)
  
\bibitem[\protect\citeauthoryear{Vazdekis et al.}{2003}]{vazdekis03} Vazdekis
  A., Cenarro A.~J., Gorgas J., Cardiel N., Peletier R.~F., 2003, MNRAS, 340,
  1317
  
\bibitem[\protect\citeauthoryear{Vazdekis, Trujillo \&
    Yamada}{2004}]{vazdekisetal04} Vazdekis A., Trujillo I., Yamada Y., 2004,
  ApJ, 601, L33
  
\bibitem[\protect\citeauthoryear{White \& Rees}{1978}]{whiterees78} White
  S.~D.~M., Rees M.~J., 1978, MNRAS, 183, 341
  
\bibitem[\protect\citeauthoryear{White \& Frenk}{1991}]{whitefrenk91} White
  S.~D.~M., Frenk C.~S., 1991, ApJ, 379, 52
  
\bibitem[\protect\citeauthoryear{Worthey, Faber \&
    Gonzalez}{1992}]{wortheyetal92} Worthey G., Faber S.~M., Gonzalez J.~J.,
  1992, ApJ, 398, 69

\bibitem[\protect\citeauthoryear{Worthey}{2004}]{worthey04} 
Worthey G., 2004, AJ, 128, 2826

\bibitem[\protect\citeauthoryear{Worthey et al.}{1994}]{wortheyetal94} Worthey
  G., Faber S.~M., Gonzalez J.~J., Burstein D., 1994, ApJS, 94, 687
  
\bibitem[\protect\citeauthoryear{Worthey \&
    Ottaviani}{1997}]{wortheyottaviani97} Worthey G., Ottaviani D.~L., 1997,
  ApJS, 111, 377
  
\bibitem[\protect\citeauthoryear{Yamada et al.}{2006}]{yamadaetal06} Yamada
  Y., Arimoto N., Vazdekis A., Peletier R.~F., 2006, ApJ, 637, 200

\end{thebibliography}
\end{document}